\documentclass[twocolumn,superscriptaddress,nofootinbib,floatfix,aps,prb,citeautoscript,longbibliography]{revtex4-2}
\usepackage[utf8]{inputenc}
\usepackage[T1]{fontenc}
\usepackage{lineno}

\usepackage{hyperref}
\usepackage{graphicx}
\usepackage{color}
\usepackage{array}
\usepackage{dcolumn}
\usepackage{bm}
\usepackage{multirow}
\usepackage[version=4]{mhchem}
\usepackage{cleveref}
\usepackage{tikz}
\usepackage{siunitx}
\usepackage{booktabs}

\newcommand{\etal}{\emph{et~al.}}
\newcommand{\meVatom}{~meV/atom}

\begin{document}

\newcommand{\halle}{Institut f\"ur Physik, Martin-Luther-Universit\"at
  Halle-Wittenberg, D-06099 Halle, Germany}
\newcommand{\coimbra}{CFisUC, Department of Physics, University of Coimbra, Rua Larga, 3004-516 Coimbra, Portugal}
\newcommand{\jena}{Institut f\"ur Festk\"orpertheorie und -optik,
  Friedrich-Schiller-Universit\"at Jena, Max-Wien-Platz 1, 07743 Jena, Germany}
  
\author{Jonathan Schmidt}
\affiliation{\halle} 
\author{Noah Hoffmann}
\affiliation{\halle} 
\author{Hai-Chen Wang}
\affiliation{\halle}
\author{Pedro Borlido}
\affiliation{\coimbra}
\author{Pedro J. M. A. Carri\c co}
\affiliation{\coimbra}
\author{Tiago F. T. Cerqueira}
\affiliation{\coimbra}
\author{Silvana Botti}
\email{silvana.botti@uni-jena.de}
\affiliation{\jena}
\author{Miguel A. L. Marques} 
\email{miguel.marques@physik.uni-halle.de}
\affiliation{\halle} 

\date{\today}

\title{Large-scale machine-learning-assisted exploration of the whole materials space}

\begin{abstract}
Crystal-graph attention networks have emerged recently as remarkable tools for the prediction of thermodynamic stability and materials properties from unrelaxed crystal structures. Previous networks trained on two million materials exhibited, however, strong biases originating from underrepresented chemical elements and structural prototypes in the available data. We tackled this issue computing additional data to provide better balance across both chemical and crystal-symmetry space. Crystal-graph networks trained with this new data show unprecedented generalization accuracy, and allow for reliable, accelerated exploration of the whole space of inorganic compounds. We applied this universal network to perform machine-learning assisted high-throughput materials searches including 2500 binary and ternary structure prototypes and spanning about 1~billion compounds. After validation using density-functional theory, we uncover in total 19512 additional materials on the convex hull of thermodynamic stability and $\sim$150\,000 compounds with a distance of less than 50~meV/atom from the hull. Combining again machine learning and ab-initio methods, we finally evaluate the discovered materials for applications as superconductors, superhard materials, and we look for candidates with large gap deformation potentials, finding several compounds with extreme values of these properties.
\end{abstract}

\maketitle

\section{Introduction}

One of the most tantalizing possibilities of modern computational materials science is the prediction and characterization of experimentally unknown compounds. In fact, developments in theory and algorithms in the past decades allowed for the systematic exploration of a chemical space spanning millions of materials, searching for compounds with tailored properties for specific technological applications. Currently, the most efficient approach consists in scanning the composition space for a given crystal structure prototype. In such approaches, the key material property that is used to estimate if a material can be experimentally synthesized is the total energy, or more specifically the energy distance to the convex hull of thermodynamic stability. Typically, for each combination of chemical composition and crystal-structure prototype one performs a geometry optimization, e.g., using some flavor of density functional theory (DFT), and compares the resulting DFT energy with all possible decomposition channels. Compounds on the convex hull (or close to it) are then selected for characterization and, if they possess interesting physical or chemical properties, proposed for experimental synthesis.

For binary prototypes this approach is relatively straightforward, and therefore the binary phase space has been comprehensively explored~\cite{aflowlib}. For a ternary prototype the different combination of chemical elements generates roughly 500\,000 compositions, a number still within reach of DFT calculations, at least for prototypes with a high symmetry and relatively few atoms in the unit cell~\cite{schmidt2017}. However, there are thousands of known ternary prototypes, making a brute force approach to the problem unrealistic. Despite the resulting huge number of candidate ternary compounds, it is worth observing that the largest computational databases only contain overall  about $4\times10^6$ materials~\cite{aflowlib,draxl2018nomad,materialsproject}. 

Machine learning methods have made it possible to accelerate material searches considerably. These methods are some of the most useful instruments added to the toolbox of material science and solid-state physics in the last decade. Thanks to them, a wide variety of material properties can now be efficiently predicted with close to ab-initio accuracy~\cite{ourreview, roadmap, Rodrigues2021}.
Early works in this direction achieved speedups by factors of about 5-30~\cite{jonathan2018, schmidt2017, 60Voronoitessellations}. These works were generally based on relatively simple machine learning models, e.g., decision trees or kernel ridge regression, that used hand-built features of the composition as input and had to be retrained for different crystal prototypes. Significant progress toward more general models was made by Ward \etal~\cite{60Voronoitessellations} who included structural descriptors applicable to high-throughput searches in terms of Voronoi tesselations. This allowed Ward \etal\ to use training data from all prototypes, resulting in improved performance for high-throughput searches. Other important steps forward were achieved by two other classes of models that were developed simultaneously: message passing networks for crystal and molecular graphs, as well as deeper composition-based models~\cite{ElemNet, goodall2019predicting, CNN2D, CNN3Dinput}. We note that compositional models can be completely independent of the crystal structure. However, they are inadequate for large-scale high-throughput searches, as they cannot differentiate between polymorphs with the same chemical composition. Message passing networks, on the other hand, enabled unprecedented performance for the prediction of  properties with \textit{ab-initio} accuracy~\cite{ourreview, dunn2020benchmarking, bartel2020critical} from crystal structures.

Until recently all message passing networks for crystals used atomic positions in some form as input. However, this information is not available until calculations, e.g. using DFT structure optimization, are performed. Consequently, such models are unpractical for high-throughput searches. Recently, Schmidt \etal~\cite{CGAT} and Goodall \etal~\cite{goodall2021rapid} developed coarse-grained message passing networks that circumvent the problem, as they do not require atomic positions as an input. In this work we apply the former approach to explore a significantly enlarged space of crystalline compounds. 

Currently, the largest issue concerning the accuracy of message passing networks is no longer the topology of the networks, nor its complexity, but is related to the limitations of existing materials datasets. Reference~\onlinecite{CGAT} identified large biases stemming from the lack of structural and chemical diversity in the available data. These biases, ultimately of anthropogenic nature~\cite{doi:10.1021/jacs.1c12005,D2DD00030J}, lead unfortunately to a poor generalization error. In fact, even if the error in test sets is of the order of 20--30~meV/atom, the actual error during high-throughput searches can be easily one order of magnitude larger if the available training data is not representative of the actual material space~\onlinecite{CGAT}.

In this work we tackle this challenging problem using a stepwise approach. First, we perform a series of high-throughput searches with an extended set of chemical elements (including lanthanides and some actinide elements), applying the transfer learning approach presented in Ref.~\cite{CGAT}. Thanks to the additional data generated by these calculations, we expect to reduce the bias due to the representation of the chemical elements in the dataset. In a subsequent step, we retrain the crystal-graph network and employ it to scan a material space of almost 1~billion compounds that  comprises more than 2000 crystal-structure prototypes. We obtain in this way a dataset of DFT calculations with a considerably larger structural diversity, that we then use to retrain a network. This crystal graph network is then shown to possess a massively improved generalization error and a strongly reduced chemical and structural bias. Finally, we offer a demonstration of the usefulness of our approach, and inspect this dataset to search for materials with extreme values of some interesting physical properties.

\section{Construction of datasets and networks}

\subsection{Enlarging the chemical space}

Our starting point is the dataset used by some of us for training in Ref.~\onlinecite{CGAT}. We will refer to this dataset as ``DCGAT-1'' and to the crystal-graph network of Ref.~\onlinecite{CGAT} as ``CGAT-1'', respectively.

As discussed previously, the training data in DCGAT-1 is biased with respect to the distribution of chemical elements and crystal symmetries. To circumvent the first problem we performed a series of high-throughput calculations for specific structure prototypes. We used a larger chemical space than previous works, considering 84 chemical elements, including all elements up to Pu (with the exception of Po and At, for which we do not have pseudopotentials, Yb whose pseudopotential exhibits numerical problems, and rare gases). This results in 6\,972 possible permutations per binary, 571\,704 permutations per ternary, and 46\,308\,024 permutations per quaternary system. For all these compositions we considered a (largely arbitrary) selection of prototypes, including ternary garnets, Ruddlesden–Popper layered perovskites, cubic Laves phases, ternary and quaternary Heuslers, auricuprides, etc. In total we included 11 binary, 8 ternaries and 1 quaternary prototypes (a complete list and more details can be found in the Supporting Information).

For each structural prototype included in the selection, we performed a high-throughput study using the transfer learning approach proposed in Ref.~\onlinecite{CGAT}: (i)~The machine-learning model is used to predict the distance to the convex hull of stability for possible chemical compositions. At the start we use the pre-trained CGAT-1 machine; (ii)~We perform DFT geometry optimizations to validate all compounds predicted below 200~meV/atom from the hull; (iii)~We add these calculations to a dataset containing all DFT calculations for the corresponding structural prototype. (iv)~We use transfer learning to train a new model on the basis of this dataset with a training/validation/testing split of 80\%/10\%/10\%. (v)~The cycle is restarted one to three times, until the MAE of the model is below 30~meV/atom.

This procedure resulted in 397\,438 additional DFT calculations, yielding 4382 compounds below the hull of DCGAT-1 (and therefore already increasing the known convex hull by approximately ten percent). Moreover, we added a large dataset of mixed perovskites~\cite{CGAT} plus data concerning oxynitride, oxyfluoride and nitrofluoride perovskites from Ref.~\onlinecite{wang2021high}, amounting to around 381\,000 DFT calculations. Finally, we recalculated and added 1343 compounds that were possibly unconverged outliers from AFLOW\cite{aflowlib} according to the criteria in Ref.~\onlinecite{aflowchull}. The final dataset resulting from all these changes and additions, contains $\sim$780\,000 compounds more than DCGAT-1 and will be denoted as DCGAT-2. 

\begin{figure}[ht!]
    \centering
    \includegraphics[width=9cm]{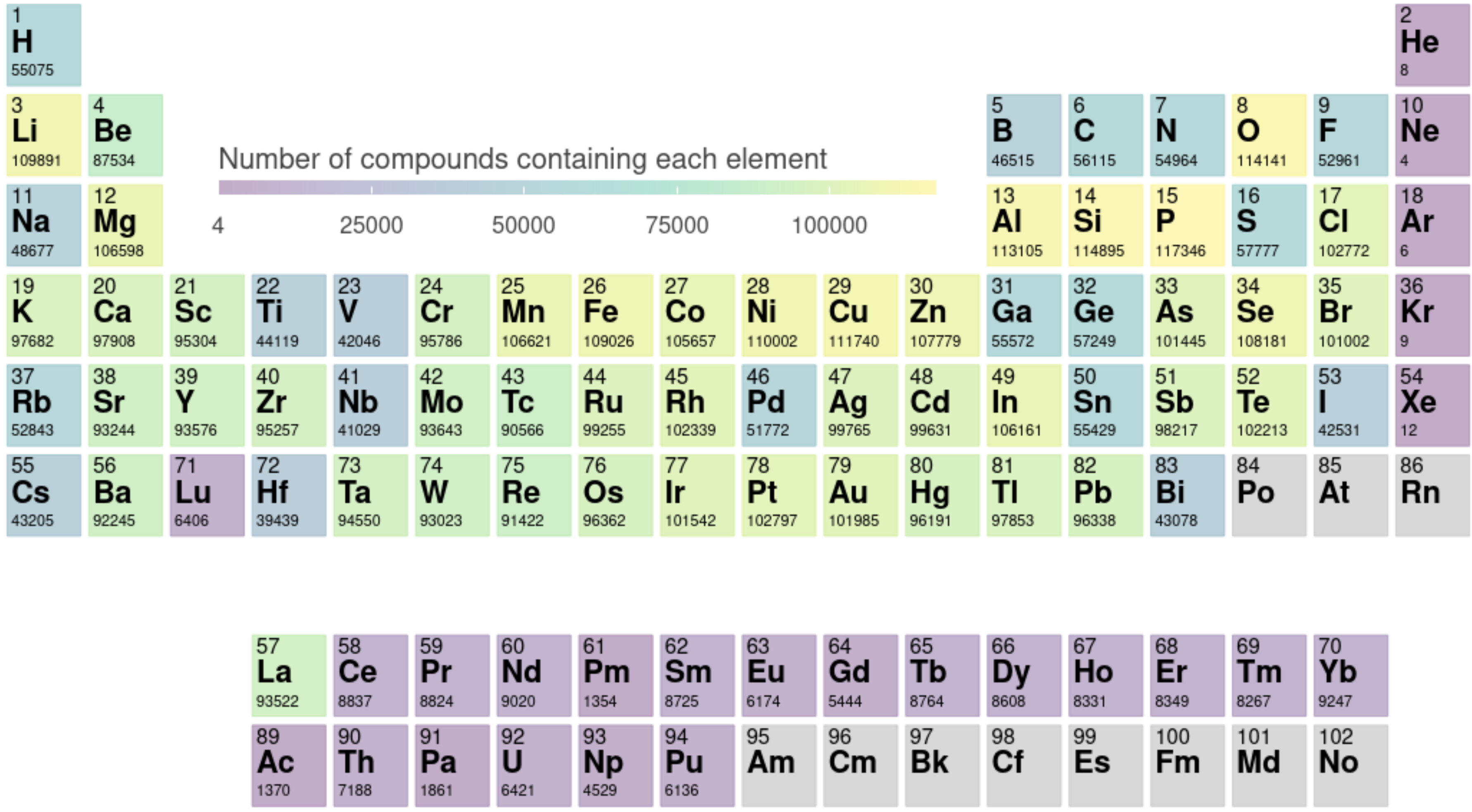}

    (a)
    
    \includegraphics[width=9cm]{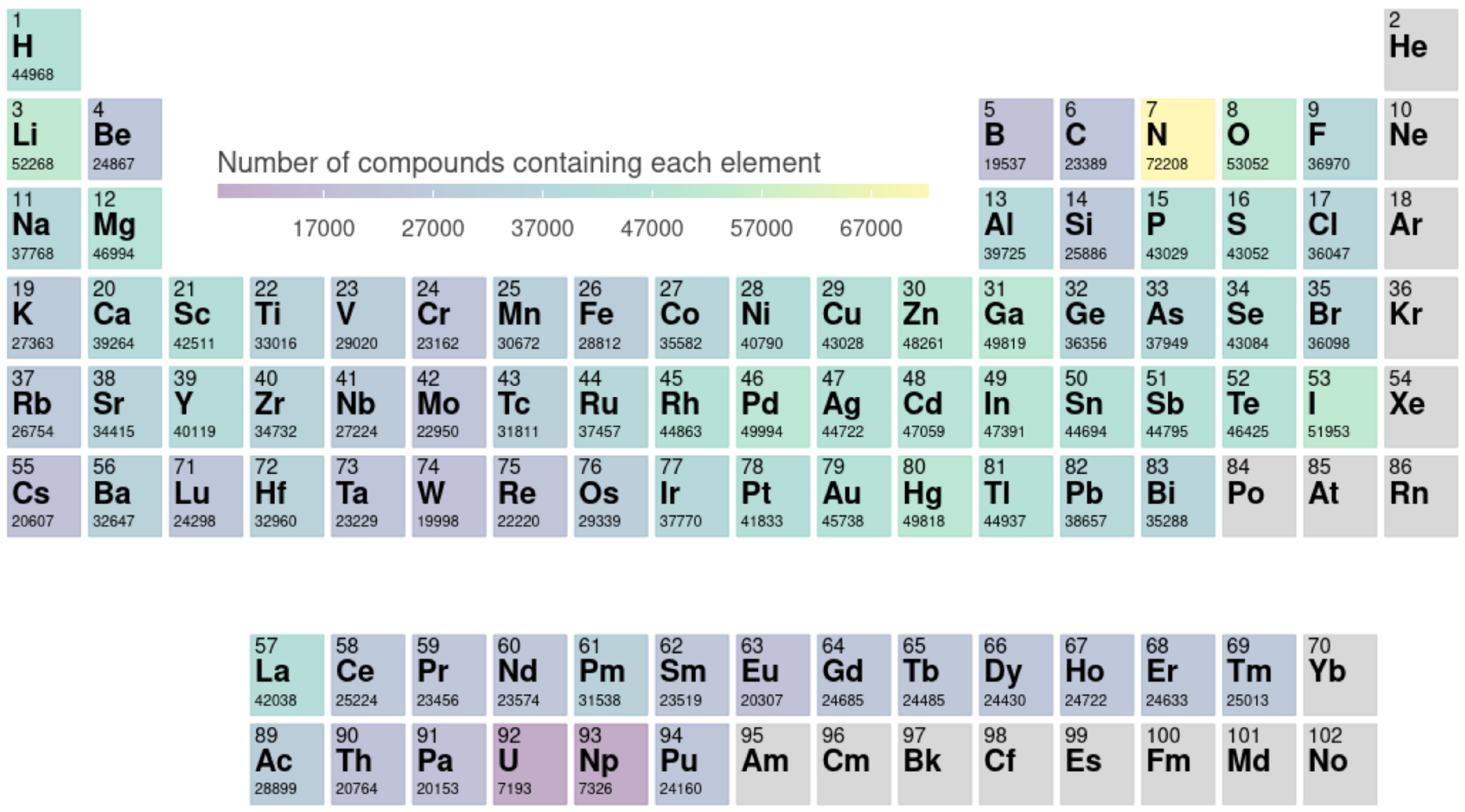}

    (b)
    \caption{Number of materials in (a)~DCGAT-1 and (b)~DCGAT-2 containing one specific chemical element of the  periodic table.}
    \label{fig:elementcount}
\end{figure}

In \cref{fig:elementcount} we plot the element distribution in both datasets DCGAT-1 and DCGAT-2. As expected, the original dataset is quite biased with a drastic undersampling of most lanthanides and actinides. Despite its smaller size, the new dataset includes between three and twenty times more compounds containing undersampled elements, and it therefore counteracts the unbalanced distribution of chemical elements of DCGAT-1. Note that, in particular, metallic elements appear in very similar quantities in the revised dataset, with exception of the heavier actinides that are still somewhat underrepresented.

We used DCGAT-2 to retrain a CGAT with the same hyperparameters used in Ref.~\onlinecite{CGAT} (the resulting network will be denoted as CGAT-2). The CGAT-2 network has a mean absolute test error of 21~\meVatom\ for the distance to the convex hull using a training/validation/test split of 80\%/10\%/10\%. Although the test error is of the same order of magnitude than CGAT-1, we will see that the generalization error is drastically reduced. We also trained a network to predict the volume per atom of the crystals, obtaining a test error of 0.25~\AA$^3$/atom. 

\subsection{Enlarging the structural space}

\begin{figure}[htb]
  \centering
  \begin{tikzpicture}[x=6cm,y=6cm]
\draw[gray] (0.9, 0.0) -- (0.45, 0.7794228634059948);
\draw[gray] (0.1, 0.0) -- (0.55, 0.7794228634059948);
\draw[gray] (0.05, 0.08660254037844387) -- (0.9500000000000001, 0.08660254037844387);
\draw[gray] (0.8, 0.0) -- (0.4, 0.6928203230275509);
\draw[gray] (0.2, 0.0) -- (0.6000000000000001, 0.6928203230275509);
\draw[gray] (0.1, 0.17320508075688773) -- (0.9, 0.17320508075688773);
\draw[gray] (0.7, 0.0) -- (0.35, 0.606217782649107);
\draw[gray] (0.30000000000000004, 0.0) -- (0.65, 0.606217782649107);
\draw[gray] (0.15000000000000002, 0.2598076211353316) -- (0.85, 0.2598076211353316);
\draw[gray] (0.6, 0.0) -- (0.3, 0.5196152422706631);
\draw[gray] (0.4, 0.0) -- (0.7, 0.5196152422706631);
\draw[gray] (0.2, 0.34641016151377546) -- (0.8, 0.34641016151377546);
\draw[gray] (0.5, 0.0) -- (0.25, 0.4330127018922193);
\draw[gray] (0.5, 0.0) -- (0.75, 0.4330127018922193);
\draw[gray] (0.25, 0.4330127018922193) -- (0.75, 0.4330127018922193);
\draw[gray] (0.4, 0.0) -- (0.2, 0.34641016151377546);
\draw[gray] (0.6, 0.0) -- (0.8, 0.34641016151377546);
\draw[gray] (0.3, 0.5196152422706631) -- (0.7, 0.5196152422706631);
\draw[gray] (0.29999999999999993, 0.0) -- (0.14999999999999997, 0.2598076211353315);
\draw[gray] (0.7000000000000001, 0.0) -- (0.8500000000000001, 0.2598076211353315);
\draw[gray] (0.35000000000000003, 0.6062177826491071) -- (0.6499999999999999, 0.6062177826491071);
\draw[gray] (0.19999999999999996, 0.0) -- (0.09999999999999998, 0.17320508075688767);
\draw[gray] (0.8, 0.0) -- (0.9, 0.17320508075688767);
\draw[gray] (0.4, 0.6928203230275509) -- (0.6, 0.6928203230275509);
\draw[gray] (0.09999999999999998, 0.0) -- (0.04999999999999999, 0.08660254037844384);
\draw[gray] (0.9, 0.0) -- (0.95, 0.08660254037844384);
\draw[gray] (0.45, 0.7794228634059948) -- (0.55, 0.7794228634059948);
\draw[gray] (0.0, 0.0) -- (0.0, 0.0);
\draw[gray] (1.0, 0.0) -- (1.0, 0.0);
\draw[gray] (0.5, 0.8660254037844386) -- (0.5, 0.8660254037844386);
    \input{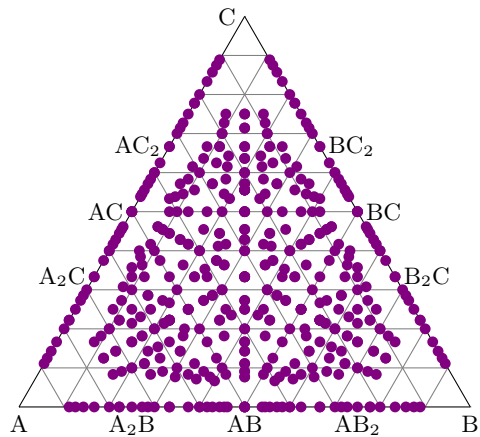}
  \end{tikzpicture}
  \caption{Ternary phase diagram showing the stoichiometries covered by the prototypes studied in this work.}
  \label{fig:pd_comp}
\end{figure}

After having successfully removed the bias in our dataset in the distribution of chemical elements, we now tackle the lack of structural variety. Our strategy consists in adding calculations of underrepresented structural types, keeping in mind that we are mainly interested in phases that are thermodynamically stable, or close to stability. We start by querying our database using the pymatgen~\cite{pymatgen2013CompMatSci} structure matcher to identify all distinct prototypes present in DCGAT-1. Note that our definition of a crystal structure prototype is relatively loose and some of them can be transformed into others through minor distortions. It is nevertheless important to keep track of all these related structures in order to increase the precision of the crystal-graph network predictions~\cite{CGAT}. We found a total of $\sim$58000 prototypes, the large majority of them appearing only once or twice in the dataset. We then selected all prototypes with less than 21 atoms in the unit cell, space-group number larger than 9 and that appeared at least 10 times in our dataset. The first two criteria are chosen to limit the run-time of the DFT calculations. Following these criteria, we end up with 639 binary and 1829 ternary crystal-structure prototypes, spanning a space of 1\,050\,101\,724 possible compounds. These prototypes also densely cover the composition space, as depicted in the generic phase diagram of \cref{fig:pd_comp}.

\begin{figure}[ht!]
    \centering
    \includegraphics[width=8cm]{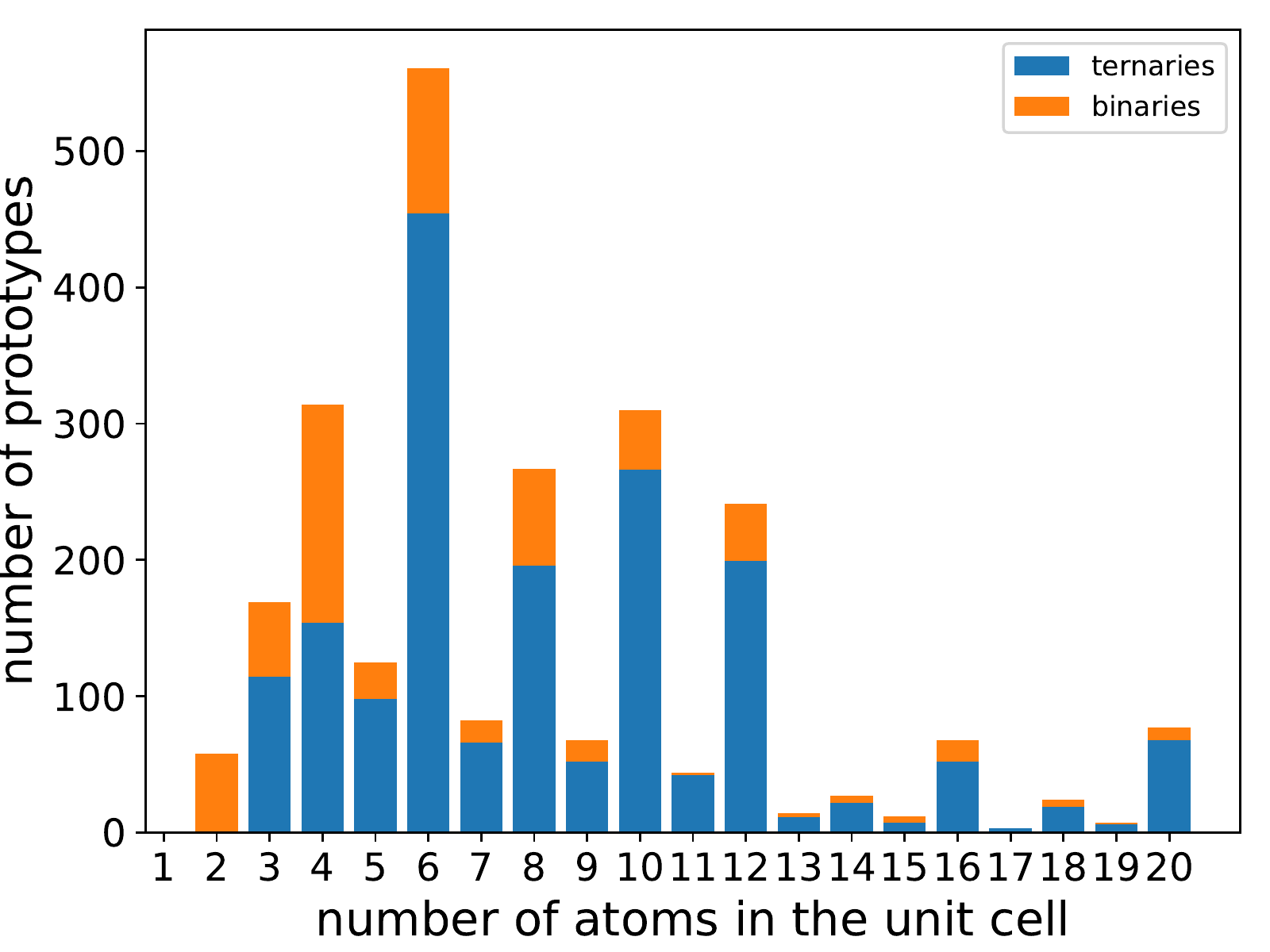}

    (a)
    
    \includegraphics[width=8cm]{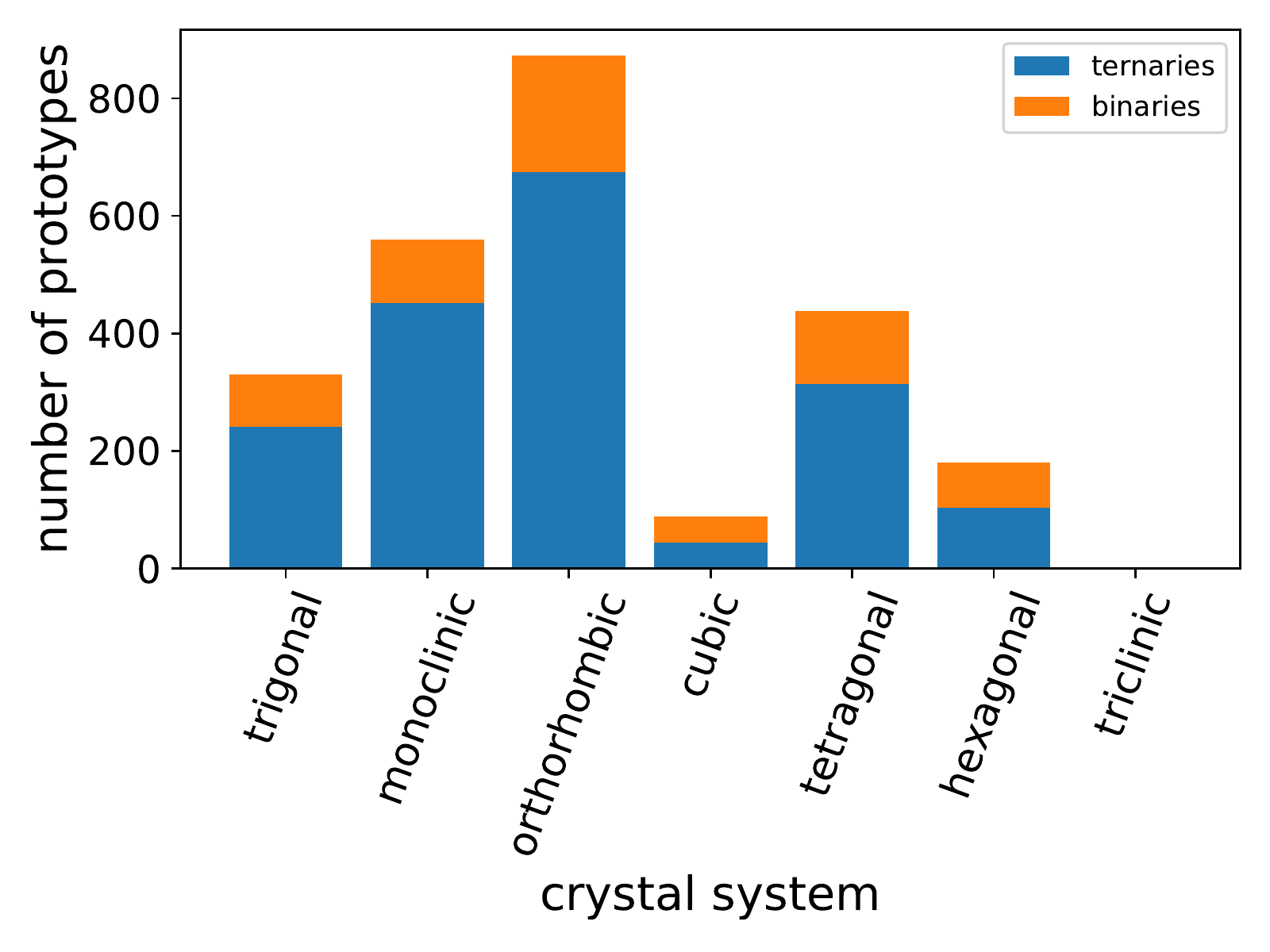}
    
    (b)
    \caption{The histograms show the distribution of (a)~number of atoms per unit cell and (b)~crystal systems in the set of prototypes scanned in the high-throughput search. The counts for the binary and ternary prototypes are stacked on top of each other in orange and blue, respectively.}
    \label{fig:atoms}
\end{figure}

In \cref{fig:atoms} we plot the distributions of number of atoms in the unit cell (panel a) and of crystal systems (panel b) in the set of selected prototypes. The distribution of the number of prototypes displays a maximum at 6 atoms per unit cell, decreasing then slowly for larger number of atoms. It is also clear that prototypes with an even number of atoms are far more common than those with odd number of atoms. The most represented crystal system is  orthorhombic, followed by monoclinic and tetragonal, while cubic prototypes are rare. Note that the number of monoclinic structures is reduced by the imposed restriction on the space group number, as monoclinic structures have space groups between 3 and 15. Also due to this restriction, no triclinic structures are present in the dataset. All these conclusions apply to both binary and ternary prototypes.

We use our CGAT-2 network to predict the distance to the convex hull for these prototypes, after grouping them according to their general composition A$_x$B$_y$C$_z$. For every composition, we occupy the lattice sites of each prototype with all permutations of the A, B, and C chemical elements, and let the machine predict the ones that are below 50~\meVatom\ above the convex hull. In case several prototypes are below this threshold we just keep the one with the lowest energy. We also remove duplicates, and materials with Eu, Gd, Yb, and Lu due to converge issues with the DFT calculations. In total we obtain 530\,937 materials below our cutoff.

\begin{figure}[ht]
    \centering
    \includegraphics[height=5cm]{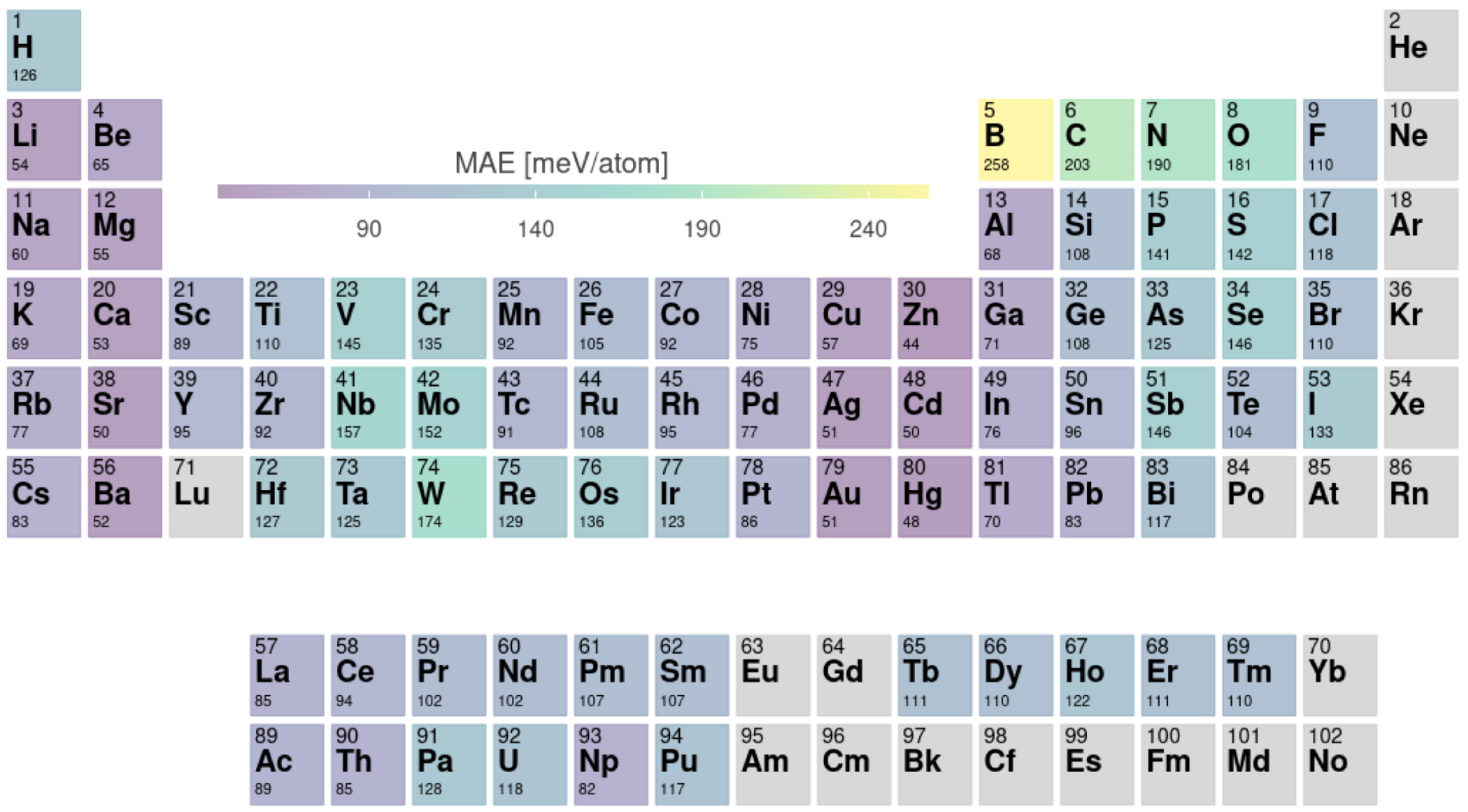}

    (a)
    
    \includegraphics[height=5cm]{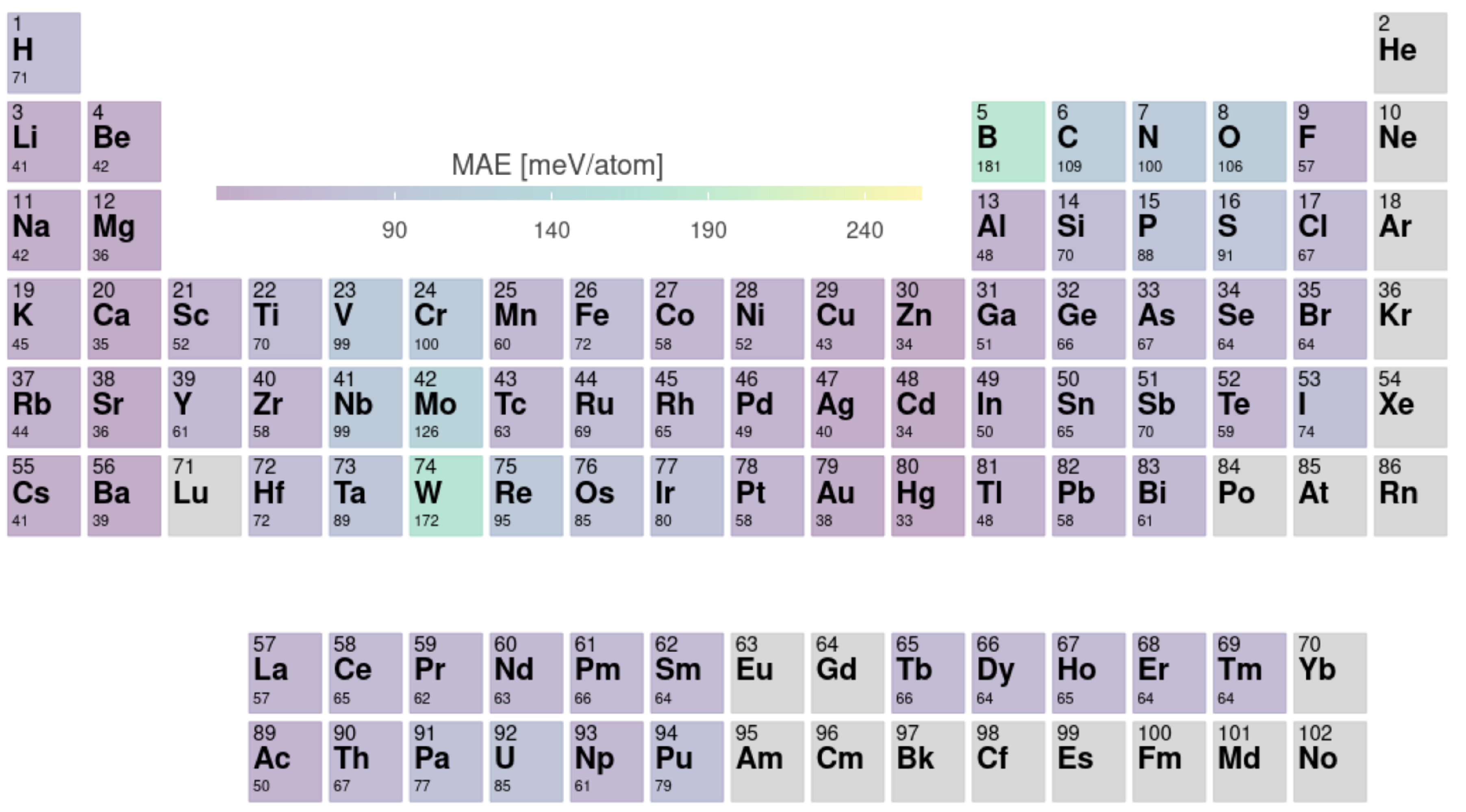}

    (b)
    \caption{MAE in the test set separated from DCGAT-3 for compounds containing each element of the periodic table, when the predictions are obtained with CGAT-1 (a) and CGAT-3 (2).}
    \label{fig:element_errors}
\end{figure}

We note the prototype geometries have not been optimized yet. We can obtain a good estimate of the unit cell volume using a CGAT network that we have trained for this quantity. We use this information to build the starting point for DFT geometry optimizations, as described in \cref{sec:methods}. After removing unconverged calculations we are left with DFT calculations for 515\,653 compounds. From this final dataset, that we call ``DCGAT-3'', we separate a testset composed of materials that correspond to 8 randomly chosen ternary compositions, encompassing 93 crystal structure prototypes and 57\,252 entries. The remaining data is then used to train our last network, called CGAT-3.

By separating a testset of compositions and prototypes sparsely represented in our training sets we expect to have a proper statistical estimation of the generalization error of the networks. The MAEs improve from 92~\meVatom\ for CGAT-1, to 87~\meVatom\,  and 57~\meVatom \,  for CGAT-2 and CGAT-3 respectively. As we can see in \cref{fig:elementcount} the element coverage in the training set already significantly improved from CGAT-1 to CGAT-2 while the prototype diversity only improved marginally. As a result the decrease in MAE is rather small at 5\%. On the other hand, the increase in prototype diversity from CGAT-2 to CGAT-3 results in a major 33\% improvement. Consequently, we can conjecture that the majority of future improvements with respect to data will come from the sampling of additional prototypes.

In \cref{fig:element_errors} we see the element resolved MAEs for CGAT-1 and CGAT-3. For CGAT-1 we observe a strong dependence of the MAE on the chemical element, with a significantly higher MAE for the first row elements, most likely due to the first row anomaly that has been observed in multiple studies~\cite{CGAT,schmidt2017,jonathan2018}. This effect is strongly reduced for CGAT-3, with a MAE that is much more uniform across the periodic table. Indeed, the maximum MAE for CGAT-1 is 258~\meVatom\ for B, while this number is reduced to to 181~\meVatom\ for CGAT-3, proving that we could essentially eliminate the chemical element bias from our dataset.

\begin{figure}[ht!]
    \centering
    \includegraphics[width=8cm]{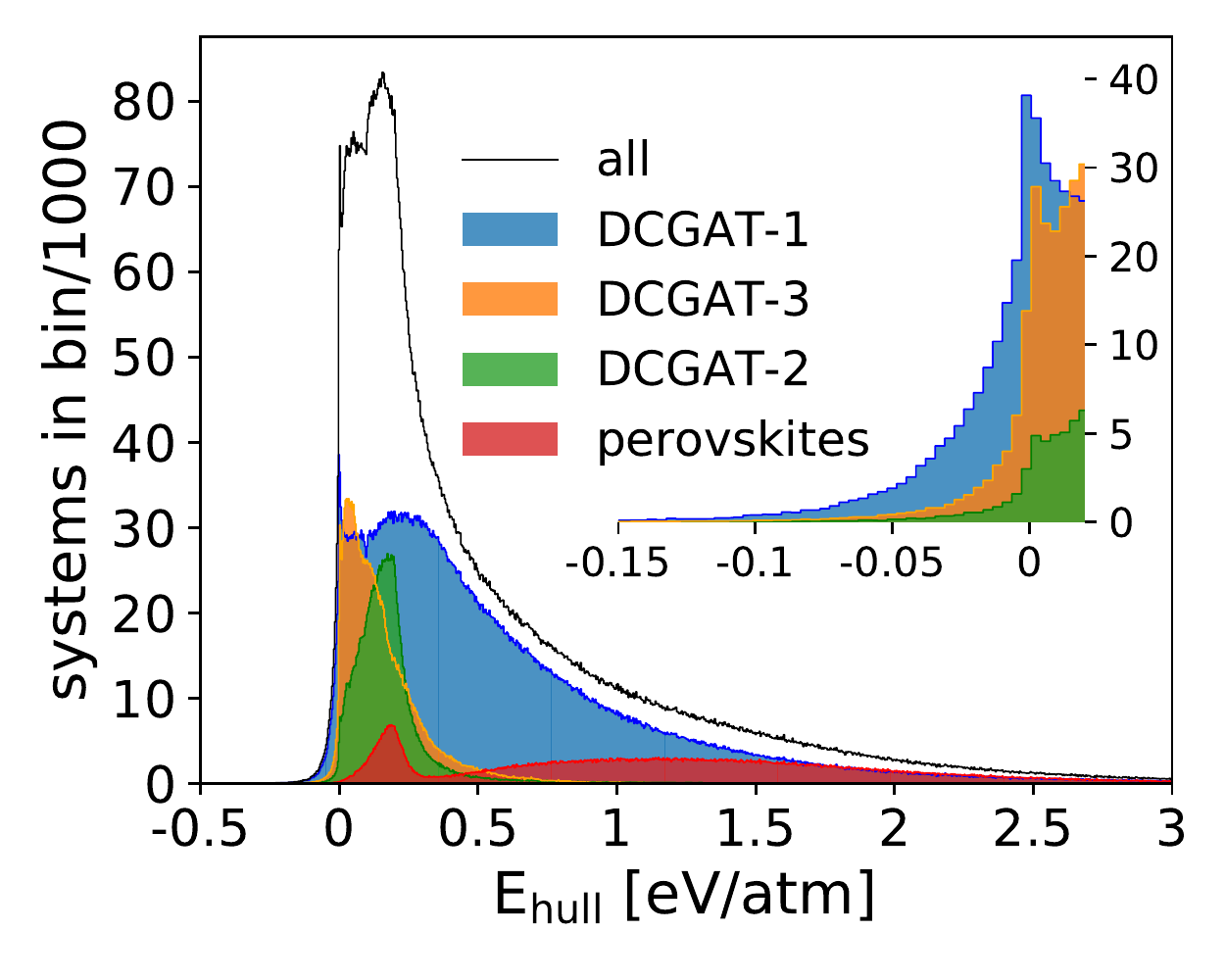}
    \caption{Distance to the convex hull for DCGAT-1 and for the data added in DCGAT-2 and DCGAT-3. The mixed perovskites studied in Ref.~\cite{CGAT} is separated from DCGAT-2 and the rest of the data. In the inset plot we zoom into the range of stable compounds.}
    \label{fig:ehulls}
\end{figure}

In \cref{fig:ehulls} we plot the distance to the convex hull for the DCGAT-1 dataset and the data that was added in DCGAT-2 and DCGAT-3. We see that the original dataset still exhibits a wide distribution with a median of 420\meVatom\ and a standard deviation of 570\meVatom. This is easy to understand as the data mostly originates from traditional high-throughput searches. The peak close to zero is due to the experimentally known stable materials as well as to the data from some studies using machine learning or chemical substitution strategies~\cite{jonathan2018,haichen}. The perovskite data added to DCGAT-2 has one peak at roughly 200~\meVatom, due the compounds generated using machine learning with a cutoff of 200~\meVatom\ from the convex hull~\cite{CGAT}. The wide distribution comes from the random perovskites generated in the same study~\cite{CGAT}. The remainder of CGAT-2 was also generated using a similar approach, and therefore is centered at 200~\meVatom. Finally, we can see that the distribution of the DCGAT-3 entries has a median of 117~\meVatom\ with a standard deviation of 154~\meVatom. Comparing to the usual range of formation energies of a high-throughput search, this distribution is extremely narrow, showing the remarkable accuracy and generalization error of our machine learning models.

\section{Material properties}
\label{sec:properties}

\begin{figure}[ht!]
    \centering
    \includegraphics[width=13cm]{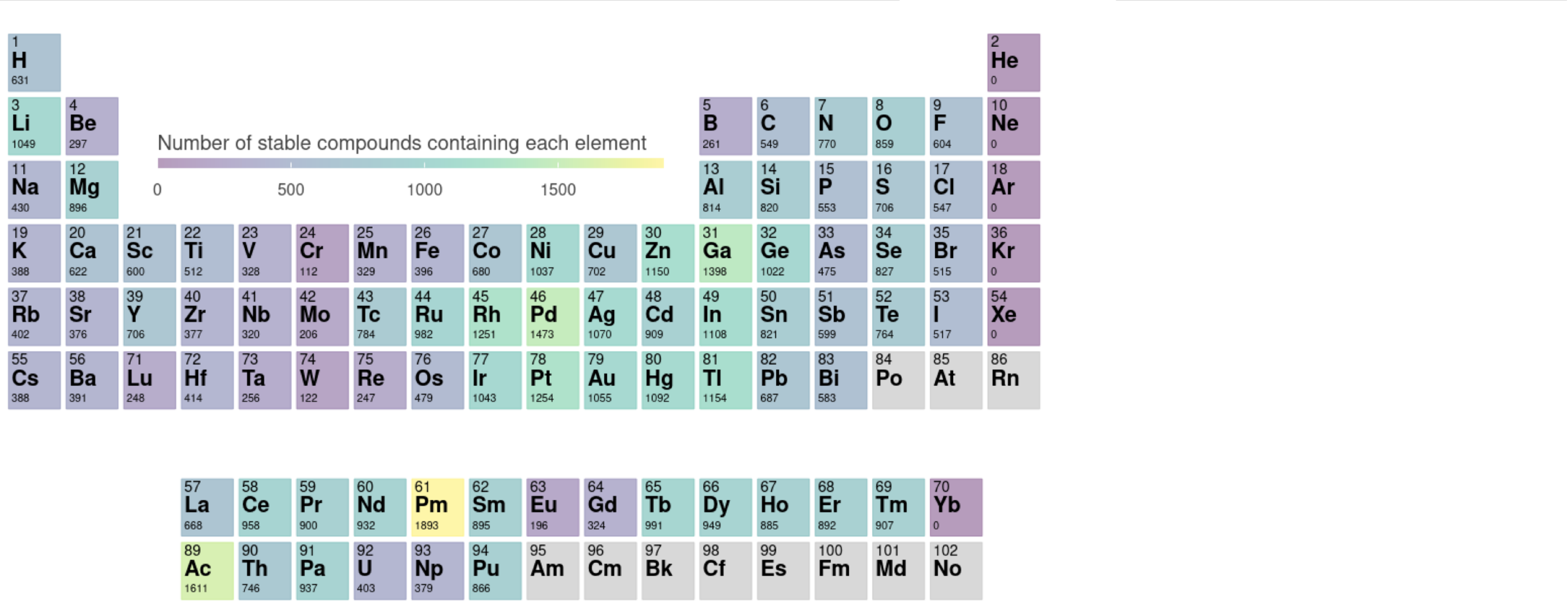}
    \caption{Number of stable compounds containing each chemical element discovered in this work.}
    \label{fig:element_count_new}
\end{figure}

In the process of constructing our unbiased datasets we have discovered 19\,512 compounds on the convex hull, 168\,340 at a distance of less than 50~\meVatom from the hull, and 326\,433 below 100~\meVatom. These crystalline materials are, to our knowledge, not yet included in available databases. An overview on the chemical nature of the new compounds on the convex hull is summarized in \cref{fig:element_count_new}, were we plot the number of newly discovered stable compounds containing each chemical element. We see that these materials cover the entire periodic table, but with a maximum for compounds including Li, Mg, transition metals around Pd and Ga, and lanthanides and actinides. Concerning the latter, we see that compounds including Eu, Gd, U, and Np are relatively underrepresented. This is not due to a lesser ability of these chemical elements to form stable compounds, but to a technical reason: the available pseudopotentials for these elements often lead to numerical problems making calculations hard to converge.

In the following we want to analyze these materials in more detail. To this end, we perform machine-learning assisted data-mining of several non-trivial physical properties to reveal compounds with extreme behavior. We decided to restrict our search to (quasi-)stable materials, defining a threshold of 50~meV/atom above the convex hull of stability. For these systems we evaluate elastic constants, superconductivity, and gap deformation potentials. The strategy in the three cases is similar: we train machine learning models based on crystal graph networks~\cite{68crystalgraphconvolution} to provide an efficient prediction of the specific property. Promising materials are then investigated in more detail using density-functional theory or density-functional perturbation theory to validate machine learning predictions and to provide further insights in the physics and the mechanism behind the extreme values of a certain property.

\subsection{Ultra-hard and incompressible materials}

\begin{table}[h]
    \centering
\begin{tabular}{lrrrrrrr}
Formula & E$_\text{hull}$ & Spg. &  N & K$_\text{VRH}$ & G$_\text{VRH}$ & H$_\text{V}$ \\ 
   \midrule
  \ce{TiVB3} &         16 &  63 & 10 &            262 &            245 &           42 \\ 
 \ce{TaTiB3} &         14 &  63 & 10 &            268 &            230 &           36 \\ 
\ce{Ta2BeB3} &          0 &  69 & 12 &            272 &            232 &           36 \\ 
\ce{BeNb2B3} &          0 &  69 & 12 &            255 &            222 &           35 \\ 
  \ce{TiB3W} &          0 &  63 & 10 &            293 &            235 &           33 \\ 
   \midrule
 \ce{Ir3Os5} &         28 &  25 &  8 &            378 &            196 &           25 \\ 
  \ce{Os2Ru} &         14 &  15 &  6 &            369 &            233 &           27 \\ 
 \ce{Os5Ru3} &         18 &  25 &  8 &            365 &            236 &           27 \\ 
\ce{MoOs4Ru} &         21 &  13 & 12 &            357 &            220 &           25 \\ 
 \ce{Os2RuW} &         49 &  51 &  8 &            350 &            189 &           23 \\ 
   \midrule
 \ce{TiVB3} &         16 &  63 & 10 &            262 &            245 &           42 \\ 
 \ce{Os5Ru3} &         18 &  25 &  8 &            365 &            236 &           27 \\ 
  \ce{TiB3W} &          0 &  63 & 10 &            293 &            235 &           33 \\ 
  \ce{Os2Ru} &         14 &  15 &  6 &            369 &            233 &           27 \\ 
\ce{Ta2BeB3} &          0 &  69 & 12 &            272 &            232 &           36 \\ 
    \end{tabular}
    \caption{Chemical formula, distance to the convex hull (E$_{hull}$), space group number (Spg.), bulk modulus ($K_\text{VRH}$ in GPa), shear modulus ($G_\text{VRH}$ in GPa), and Vicker's hardness ($H_\text{V}$ in GPa) for the materials with the highest calculated $H_\text{V}$ (top section), $K_\text{VRH}$ (middle section) and $G_\text{VRH}$ (bottom section). All indicated materials satisfy the Born-Huang elastic stability criteria~\cite{PhysRevB.90.224104,born_1940}. }
    \label{tab:nice_elastic_materials}
\end{table}

\begin{figure}[ht!]
    \centering
    \includegraphics[width=6cm]{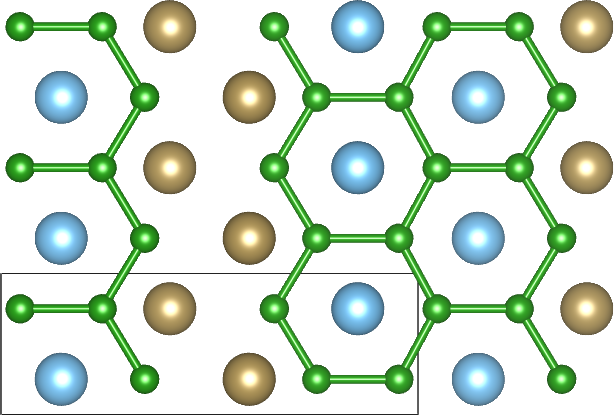}
    \caption{Crystal structure of \ce{TiTaB3}. The blue atoms represent Ti, the gold Ta, and the green B. We also depict the primitive unit cell. Picture produced with \textsc{vesta}~\cite{vesta}.}
    \label{fig:TiTaB3}
\end{figure}

Describing the elastic response of a material requires knowledge of its stiffness tensor, composed of (at most) 21 independent elastic constants. A direct analysis of the whole tensor is rather cumbersome, so most studies concentrate on two derived properties: the Voigt-Reuss-Hill averaged~\cite{0370-1298-65-5-307} bulk and shear moduli $G_\text{VRH}$ and $K_\text{VRH}$, respectively, that describe the compressibility of the material. Besides considering the average elastic response of the material, we can use $G_\text{VRH}$ and $K_\text{VRH}$ to estimate the Vicker's hardness~\cite{smith1922ProcIntsMechEng, mazhnik2019JAP} $H_\text{V}$ and use this quantity to identify ultra-hard materials.

A recent example of this approach is Ref.~\onlinecite{zuo2021accelerating}, where
the authors used Bayesian optimization with symmetry relaxation to obtain optimized structures, followed by materials graph network~\cite{megnet} models to predict formation energies as well as $G_\text{VRH}$ and $K_\text{VRH}$. This methodology was applied to search for ultra-hard transition metal borides and carbides, exploring a comparatively small space of circa 400\,000 compounds.

In this work we perform a screening of the values of $G_\text{VRH}$, $K_\text{VRH}$ and $H_\text{V}$ on our much larger dataset. We predict the values of the first two quantities using CGCNN models trained on the dataset of Matbench~\cite{dunn2020benchmarking}, and use these predicted values to estimate the Vicker's hardness for each material using the model of Ref.~\cite{mazhnik2019JAP}. The compounds within 50~meV/atom from the convex hull and with the 25 highest $G_\text{VRH}$ and $K_\text{VRH}$ were selected for subsequent analysis, i.e., their stiffness tensors were calculated using DFT, as described in \cref{sec:methods}. The top-5 materials for each quantity are presented in Table~\ref{tab:nice_elastic_materials} (while a complete list is given in the SI).

Not surprisingly, the materials with the highest $H_V$ are mostly metal borides. These materials are known to emulate the hardness of diamond thanks to a mixture of high valence (provided by the metal) and short bonds (provided by boron)~\cite{ultrahardboridesperspective2018, Pangilinan2022,tawfik2022}. Most of the borides seen here belong to the same prototype, with anonymous formula \ce{MNB3}, with M and N a metal, and space group 63. This crystal structure, depicted in \cref{fig:TiTaB3}, consists of hexagonal boron nanoribbons intercalated with layers of transition metals.
Overall, this arrangement of atoms is reminiscent of other ultra-hard materials such as \ce{WB}, \ce{WB2} and \ce{TiB2}.
We remark the prediction of the superhard ternary compound \ce{TiVB3} with a hardness of 42~GPa.

For what concerns incompressible materials, the presence of osmium compounds (\ce{Ir3Os5} and \ce{Os2Ru}, etc.) is also not too surprising, as osmium and osmium compounds are known to have extremely large bulk modulus, although they are not necessarily hard, due to the metallic nature of their bonds~\cite{superhardreview2016}.

\subsection{Superconductors}

\begin{table}[htb]
    \centering
\begin{tabular}{lrrrrr}
      Formula & $E_\text{hull}$ & Spg. & $\lambda$ & $\omega_{log}$ [K]& T$_c$ [K] \\
    \midrule
  \ce{ScMoC2} &         49 & 166 &      0.86 &            287.94 & 15.66 \\
 \ce{NbRhBe4} &         47 & 216 &      0.70 &            311.69 & 10.91 \\
  \ce{YZr3N4} &         35 & 221 &      0.54 &            398.16 &  6.56 \\
  \ce{Sc4NO3} &          8 & 221 &      0.50 &            428.04 &  5.12 \\
  \ce{Zr4CN3} &          0 & 221 &      0.50 &            419.02 &  5.05 \\
 \ce{ScZr3N4} &          0 & 221 &      0.48 &            440.18 &  4.50 \\
    \end{tabular}
    \caption{Formula, distance to the convex hull ($E_\text{hull}$), space group (Spg.), and calculated superconducting properties for the screened materials.}
    \label{tab:topsc}
\end{table}

Searching for new conventional superconductors with high critical temperature ($T_\text{c}$) is always a tempting application for large material datasets. This turns out to be a complex task due to the interplay between the different ingredients that determine $T_\text{c}$, as well as the lack of reliable simple indicators of superconductivity~\cite{Lilia_2022_supercond_roadmap}. McMillan's formula~\cite{mcmillan1968PRB} suggests to use the Debye's temperature ($\Theta_\text{D}$) and the density of states at the Fermi level (DOS(E$_F$)) as estimators for high-$T_c$, a connection that has recently been used with some success~\cite{choudhary2022arxiv}. Within the context of the present work we can easily estimate $\Theta_\text{D}$ from the bulk and shear moduli~\cite{anderson1963JPhysChemSol} obtained in the previous section.

Following this line of thought, we selected non-magnetic materials with a predicted Debye's temperature above 300\,K and with a DOS(E$_F$) larger than $0.5$~states/eV. Furthermore, we restricted our search to space-group numbers greater or equal to 160 (highly symmetric compounds), and to cells with a maximum of 8 atoms. We ordered the resulting 2717 materials by $\Theta_D$, and performed electron-phonon calculations for the first 50. The large majority of these are dynamically stable (2 were found to have imaginary frequencies), and we found 19 system with $T_c$ above 1\,K, as calculated using McMillan's formula. For the compounds with largest calculated $T_c$, we performed more converged electron-phonon calculations by increasing the density of the $q$- and $k$-grids. The calculated superconducting properties for these compounds can be seen in Table~\ref{tab:topsc}, while a complete table for the 50 screened materials is available in the SI.

\begin{figure}[ht!]
    \centering
    \includegraphics[width=9cm]{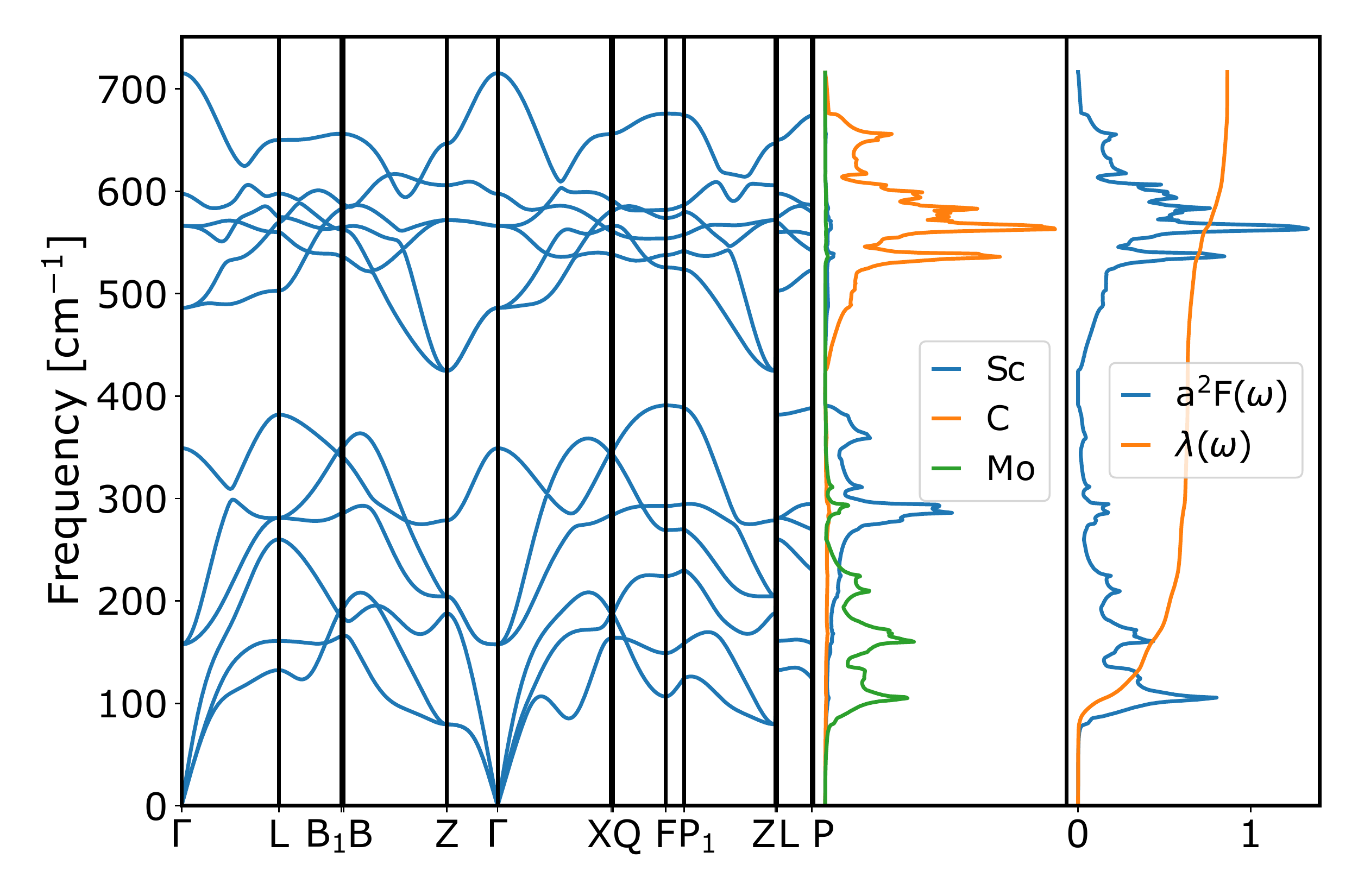}
    \caption{Phonon dispersion, density of states, and electron phonon coupling $\alpha^2F(\omega)$ of \ce{ScMoC2}.}
    \label{fig:C2ScMo_ph-bs}
\end{figure}

The compound with highest transition temperature in Table~\ref{tab:topsc} is \ce{ScMoC2}, with a $T_\text{c}=15.97$~K. This is a very interesting material with a rhombohedral lattice exhibiting alternating layers of Sc-C-Mo-C. The lowest frequency phonon modes have Mo-character, while the optical phonon modes until around 400~cm$^{-1}$ have mostly Sc-character. These are separated by a gap from a manifold of optical modes exclusively due to the C-atoms. A large part of the electron-phonon coupling constant $\lambda$ comes from a softening of an acoustic branch in the $\Gamma\rightarrow$X direction, and that is ultimately responsible for the large value of $T_\text{c}$.

The only intermetallic compound in the top 5 list is \ce{NbRhBe4}. This is actually a ternary generalization of the cubic Laves (C15) phase~\cite{10.1007/s10853-020-05509-2}. Interestingly, both \ce{NbBe2} and \ce{RhBe2} have been synthesized~\cite{10.1088/1468-6996/16/3/033503}, and are superconducting with T$_\text{c}$ of 2.14~K and 1.37~K respectively. Our prediction of 11.61~K is considerably higher than for each of the individual binaries, but in line with the related A15 compound \ce{Nb3Be} that has T$_\text{c}=10$~K~\cite{10.1134/1.1633313}.

Interestingly, the list in Table~\ref{tab:topsc} also includes several nitrides. As an example, we take a closer look at \ce{ScZr3N4} (that is isostructural to \ce{YZr3N4} also on the list). This material has a simple cubic structure (space group Pm$\bar3$m \#221) that can be derived from the NaCl-type structure, with N occupying one site (Wyckoff positions 1b and 3d), and the cations occupying the other site (Sc in the 1a  and Zr in the 3c Wyckoff positions). The phonon dispersion and density of states, and the electron-phonon coupling $\alpha^2F(\omega)$ of \ce{ScZr3N4} can be found in the SI. Consistently with the difference in atomic masses of the composing atoms, the acoustic and lower optical branches have Zr character, the following manifold just below 300~cm$^{-1}$ has mostly Sc character, and the highest manifold at around 400--500~cm$^{-1}$ is related to N. Finally, all modes contribute to the coupling constant $\lambda$ that reaches a value of around 0.8.

\subsection{Deformation Potentials}

\begin{table}
    \centering
    \begin{tabular}{l c c c c c}
Material & $E_\text{hull}$ & Spg. & $E_g$ &  $\Xi$  & $\tilde{\Xi}_\text{pred}$ \\ \midrule
\ce{InGaO3         } &      49 &    148  &   1.48  &    -7.76  &   -8.02  \\
\ce{NaLi4F5        } &      42 &    139  &   7.64  &    -7.30  &   -7.40  \\
\ce{AcAlF6         } &       0 &    166  &   7.31  &    -7.22  &   -7.30  \\
\ce{NaLi3F4        } &      43 &     65  &   7.64  &    -6.88  &   -7.03  \\
\ce{Na2SiN2        } &      14 &     72  &   2.13  &    -6.30  &   -8.53  \\
\ce{PaIO6          } &       0 &    148  &   2.01  &    -5.95  &   -8.31  \\
\ce{Li3ClF2        } &      27 &     71  &   6.35  &    -5.82  &   -6.92  \\
\ce{KNdF4          } &       9 &    123  &   6.56  &    -5.75  &   -6.28  \\
\ce{AcGaF6         } &       0 &    166  &   5.83  &    -5.68  &   -7.77  \\
\ce{LiTlSe         } &      30 &     11  &   1.01  &    -5.38  &    3.64  \\
\midrule
\ce{Pb5SeS4        } &       7 &    139  &   0.42  &     3.20  &    3.35  \\
\ce{TlIn4Cl5       } &      23 &     87  &   1.69  &     4.04  &    4.14  \\
\ce{TlIn4Br5       } &      11 &     87  &   1.49  &     4.08  &    3.61  \\
\ce{In4GaBr5       } &      31 &     79  &   1.43  &     4.08  &    3.21  \\
\ce{LiTl4I5        } &      30 &    166  &   1.94  &     4.12  &    3.18  \\
\ce{In5Br4Cl       } &      18 &    166  &   1.39  &     4.23  &    3.45  \\
\ce{In3Br2Cl       } &      22 &     44  &   1.43  &     4.61  &    3.63  \\
\ce{InHg2F         } &      36 &     11  &   0.82  &     5.26  &    4.58  \\
\ce{LiGeF3         } &       0 &    148  &   4.14  &     5.80  &    3.24  \\
\ce{TlHg2F         } &      17 &     59  &   1.23  &     6.81  &    3.82  \\
    \end{tabular} 
    \caption{Materials with the largest (in absolute value) hydrostatic deformation potentials.
    Shown here are the chemical formula, distance to the hull ($E_\text{hull}$ in meV/atom), space group number (Spg.), band gap ($E_g$ in eV), deformation potential ($\Xi$ in eV) and screened deformation potential ($\tilde{\Xi}_{screen}$,in eV). }
    \label{tab:def_pot_examples}
\end{table}

Finally we take a look at the hydrostatic deformation potentials $\Xi$, which measure the variation of the band gap ($E_g$) with respect to hydrostatic variations of the structure. This quantity is defined as
\begin{align}
    \Xi &= \frac{\mathrm{d} E_g }{\mathrm{d} \ln(V)} \nonumber \\ 
    & = \frac{ \partial E_g }{ \partial \ln(V)}
    +
    \sum_{i}
    \frac{ \partial E_g }{ \partial u_{i} } 
    \frac{ \partial u_{i} }{ \partial \ln(V) }
    \label{eq:def_pot_def}
\end{align}
where the second term of the last equation comes from the dependency on the internal parameters (i.e. atomic positions and cell vectors). For large scale studies, the use of the complete Eq.~\eqref{eq:def_pot_def} is rather cumbersome, and it is preferable to resort to the fixed shape hydrostatic deformation potential, $\tilde{\Xi} \equiv  \frac{ \partial E_g }{ \partial \ln(V)}$) to detect large deformation potentials.
Using the dataset of Ref.~\cite{defpot}, which provides $\tilde{\Xi}$ for a series of semiconductors, we trained a CGCNN model to predict this quantity (which we define by $\tilde{\Xi}_\text{pred}$) for the materials in the present dataset. We considered compounds within 50~\si{\milli\eV}/atom from the hull, with a maximum of 10 atoms in the unit cell and electronic band gaps larger than 0.1~eV. From the predicted values of $\widetilde{\Xi}$, we identified a set of 338 extreme materials for which we calculated the full deformation potential $\widetilde{\Xi}$. A summary of the system with largest absolute values of the deformation potentials can be found in Table~\ref{tab:def_pot_examples}.

Materials with extreme negative gap deformation potentials are very diverse, both in terms of their chemistry and the size of their band gap. In fact, we find oxides, fluorides, nitrides, etc., with band gaps ranging from 2.0~eV to more than 7.5~eV. We also note the appearance of alloyed systems, with two closely related cations, such as \ce{InGaO3} or {NaLi4F5}. The latter material is a very interesting example: \ce{NaLi4F5} and \ce{NaLi3F4} are ordered alloys of \ce{NaF} and \ce{LiF}. Their end components are wide gap materials (6.1~eV for \ce{NaF} and 8.7~eV for \ce{LiF}) with suitable refractive indices to be used in ultraviolet optics and Cherenkov radiators~\cite{arnoldNIaMiPRSAASDaAE1988}, but are perhaps best known for their use in molten salt reactors~\cite{frandsenJoNM2020}.
Owing to the isoelectronic substitution of Na by Li, the electronic structure of these ordered alloys is qualitatively similar to that of \ce{NaF} and \ce{LiF}: F's $p$-orbitals contribute heavily  to the valence band while the conduction band shows primarily $s$-orbitals with contributions from all elements. As expected, their band gap also lies in between that of NaF and LiF, at 7.6~eV for both entries. The values of the deformation potentials are also very similar between the two, -7.30~eV for \ce{NaLi4F5} and -6.88~eV \ce{NaLi3F4}.

A similar reasoning follows for \ce{InGaO3}, which crystalizes in the ilmenite structure (space group $R \bar{3}$), a derivative of the corundum structure family.
This material is essentially an ordered alloy of \ce{Ga2O3} and \ce{In2O3}, and along with \ce{Al2O3} it forms an isoelectronic set of corundum phases. These materials have already been observed to possess high deformation potentials, with aluminum corundum being commercially used to measure pressure inside diamond anvil cells.
As Ga and In are neighbors in the periodic table, alloying Ga and In in this structure leads to a band structure that is qualitatively very similar to that of the end components.
The valence band is dominated by contributions of oxygen's $p-$orbitals, making it very ```flat'', while the almost parabolic conduction band is more complex, showing an admixture of $s-$ orbitals from In, O and Ga, as well as $p-$orbitals from O.
If we consider the isoelectronic sequence  \{Al, Ga, In\}$_2$\ce{O3}, we observe a decrease of the (PBE) band gap with increasing atomic number (5.85, 2.40, 0.96~eV~\cite{materialsproject}, respectively) and a decrease of the absolute deformation potential (-12,-10,-8~eV~\cite{defpot}, respectively). The band gap and deformation potential of \ce{InGaO3} lies in the middle of the range of values.

These encouraging results point to the possibility of engineering the band gap deformation potentials of \ce{(Li,Na)F} or \ce{(In,Ga)O3} alloys by controlling the ratio Li/Na or In/Ga in the aforementioned phases.  Mixing these compounds is not energetically favourable and therefore we expect the formation of random alloys at adequate temperatures. A detailed study of the thermodynamics of these alloys would be necessary to make quantitative predictions.

Finally, we note the presence of a false positive, \ce{LiTlSe}, on the list of Table~\ref{tab:def_pot_examples}. In this compound the relaxation of the internal coordinates leads to an enormous correction of the fixed shape deformation potential, even leading to a change of sign. \ce{LiTlSe} belongs to the matlockite family (e.g.~\cite{hajhassanJoPaCoS2004, reshakPBCM2008}) and this crystal is almost layered. The structure is comprised of sheets of \ce{LiSe}, where the Li atoms are the center of flat tetrahedra with Se at the vertices. These are arranged in a square lattice, such that both vertices and edges are shared between adjacent tetrahedra. The Tl atoms are placed in the concavities of the tetrahedra, thus separating the \ce{LiSe} layers. The valence band shows a predominance of Se-$p$ orbitals followed by Tl-$s$ orbitals, while the conduction is primarily owed to Tl- and Se-$p$-orbitals.

Also the compounds with the highest positive gap deformation potentials display a large variety of chemistry and band gaps. The latter range from 0.4~eV to more than 4~eV. The material with largest values of $\Xi$ is \ce{TlHg2F}.

\section{Conclusions}

We propose a universal crystal-graph attention neural network that predicts the convex hull of the whole materials space with unprecedented accuracy, from the sole knowledge of prototype crystal structures. To obtain this result, we removed biases originating from underrepresented chemical elements and structural prototypes in the training dataset of materials calculations.

Applying our neural network we scrutinized nearly a billion materials and were able to expand the known theoretical convex hull by roughly 30\%, revealing tens of thousands of realistic targets for experimental synthesis. To exemplify how to take advantage of the uncovered opportunities for materials discovery, we further predicted a selection of material properties using a combination of machine learning and standard approaches. In this way we discovered a number of ultra-hard and superconducting materials, as well as materials with extreme gap-deformation potentials. We suggest with highest priority as interesting synthesis targets, e.g., ultra-hard \ce{TiVB3} or superconducting \ce{ScMoC2} with a predicted critical temperature of 16~K.

Our results point to the importance of the quality of the training data and demonstrate that creating additional and diverse data is the key to improve large-scale machine learning models in material science, so that they perform with a consistently small error across the prototype and composition space. As an extension of this work, we are currently looking over quaternary prototypes and these new calculations will soon further enlarge and diversify our materials dataset. As a perspective, with our data-driven approach we aspire in the near future to reduce the false negative rate to such an extent that machine learning predictions will largely replace DFT-based high-throughput searches.

\section{Methods}
\label{sec:methods}
\subsection{DFT Calculations}

\subsubsection{Geometry relaxations}

We performed all geometry optimizations and total energy calculations with the code VASP~\cite{vasp1,vasp2}. All parameters for the calculations were chosen to be compatible with the materials project database~\cite{materialsproject}. The Brillouin zones were sampled by uniform $\Gamma$-centered k-point grids with a density of 1000 $k$-points per reciprocal atom. We applied the projector augmented wave parameters~\cite{paw,paw2} of VASP version 5.2 with a cutoff of 520 eV. We converged the calculations to forces smaller than 0.005 eV/\AA. As exchange-correlation functional we used the Perdew-Burke-Ernzerhof~\cite{PBE} functional with on-site corrections for oxides, fluorides containing Co, Cr, Fe, Mn, Mo, Ni, V, and W. The repulsive on-site corrections to the d-states were respectively 
3.32, 3.7, 5.3, 3.9, 4.38, 6.2, 3.25, and 6.2 eV.
We encountered convergence issues with heavy elements, like Pu for which the calculations often did not converge within our time limits, and several Lanthanides, e.g., Gd and Eu for which the self-consistent cycles sometimes did not converge. Furthermore, Cs has a problematic pseudopotential which leads to additional unconverged calculations. Unconverged calculations were eliminated from the datasets.

\subsubsection{Elastic constants}

Calculation of the stiffness tensors were performed using DFT with VASP~\cite{vasp1,vasp2} via \texttt{atomate}~\cite{atomate2017CompMatSci} workflows, using the corresponding default input parameters. In a nutshell~\cite{de2015charting}, the calculation was done straining the cell with 6 deformation gradients, in 4 different magnitudes, for a total of 24 distorted cells. From the results of these calculations we fitted and suitably symmetrized the components of the stiffness tensor.
Once the stiffness tensor is known, all derived quantities ($K_\text{VRH}$, $G_\text{VRH}$, $H_\text{V}$) can be trivially obtained~\cite{mazhnik2019JAP,0370-1298-65-5-307}.

\subsubsection{Electron-phonon coupling}
We performed electron-phonon calculations using the version 7.0 of \textsc{quantum espresso}~\cite{quantumespresso} with the PBE for solids (PBEsol)~\cite{pbe-sol} generalized gradient approximation. We used pseudopotentials from the \textsc{pseudodojo} project~\cite{pseudodojo}, specifically the PBEsol stringent norm-conserving set. This pseudopotential table has been systematically constructed and validated in a series of 7 tests in crystalline environments, specifically the $\Delta$-Gauge~\cite{eessdft}, $\Delta'$-Gauge~\cite{deltagauge_prime}, GBRV-FCC, GBRV-BCC, GBRV-compound~\cite{GBRV_compound}, ghost-state detection, and phonons at the $\Gamma$-point. 

Our workflow consisted in the following steps:
(i)~The energy cutoff was set to the maximum of \textsc{pseudodojo}'s high precision hint of the elements in a given material. 
(ii)~The lattice constant was optimized using uniform $\Gamma$-centered $k$-point grids with a density of 1500 $k$-points per reciprocal atom. If this resulted in an odd number of $k$-points in a given direction, the next even number was used instead. Convergence thresholds for energies, forces and stresses were set to $1\times10^{-8}$\,a.u., $1\times10^{-6}$\,a.u., and $5\times10^{-2}$~kbar, respectively. For the electron-phonon coupling we used a double grid technique, with the same $k$-grid used in the lattice optimization as the coarse grid, and a $k$-grid quadrupled in each direction as the fine grid. 
(iv)~For the $q$-sampling of the phonons we used half of the $k$-point grid described above.
(v) The double $\delta$-integration to obtain the Eliashberg function was performed with a Methfessel–Paxton smearing of 0.03~Ry.
(vi)~The values of $\lambda$ and $\omega_\text{log}$ were then used to calculate the superconducting transition temperature using the Allen-Dynes modification~\cite{allen1975PRB} to the McMillan formula~\cite{mcmillan1968PRB}
\begin{equation}
    T_\text{c} = \frac{w_\text{log}}{1.20} \exp\left[-1.04\frac{1 + \lambda}{\lambda - \mu^*(1 + 0.62\lambda)}\right]
    \,.
\end{equation}
We took arbitrarily the value of $\mu^*=0.10$ for all materials studied. For the higher accuracy calculations, we repeated the previous steps by changing: (i) the initial $k$-point grid density used for the geometry optimization was set to 3000 $k$-points per reciprocal atom; (ii) the $k$-grid used as the coarse grid was set to the double of the $k$-grid used for the geometry optimization.

\subsubsection{Deformation Potentials}

Calculation of the deformation potentials was done within density-functional theory with VASP~\cite{vasp1,vasp2} and the PBE approximation~\cite{PBE} to the exchange correlation potential. Geometry optimizations were performed using $\Gamma$-centered grids with 1500~$k$-points per reciprocal atom until the forces were smaller than 5~\si{\milli\eV\per\angstrom}. Densities of states were calculated using grids with 2000 $k$-points per reciprocal atom and band-structure using a line density along the high-symmetry path of $60 \cdot 2\pi$~\si{\per\angstrom}. The non-spherical contributions from the gradient corrections inside the augmentation spheres were included as well. Apart from this, the remaining inputs (e.g. pseudopotential choice and Hubbard parameters) were chosen to be the same as recommended by the Materials Project. For the band structures we used the notation of Ref.~\cite{setyawan2010CompMatSci} to build the paths in reciprocal space, with the conversion to the standard representation being handled by the \texttt{pymatgen} package~\cite{pymatgen2013CompMatSci}.
The deformation potentials were computed from the band-structures calculated at three different optimized cell volumes: the optimized volume, a volume compressed by $3\%$ and a volume expanded by $3\%$. For the distorted cells we performed a geometry optimization at fixed volume. Finally, the deformation potentials were obtained by fitting a first-order polynomial to reproduce the resulting $E_g$ and $\ln(V)$ data.

\subsection{Machine Learning}

We use CGATs for the prediction of the distance to the convex hull and the volume of the crystal structures. CGATs are message passing networks on crystal graphs relying on the attention mechanism\cite{vaswani2017attention} to construct the messages and updates. We denote the vector representing the $i_{\text{th}}$ node, i.e. atom, at time step t of the message passing process as $h_i^t$ and the corresponding edge to the atom j as $e^t_{ij}$. In general, the message passing and update equation can be summarized as: 
\begin{equation}
    \mathbf{h}^{t+1}_i = U\left(\mathbf{h}^t_i,\left\{\mathbf{h}^t_j, \mathbf{e}^t_{ji}\right\},j\in \mathcal{N}(i) \right).
\end{equation}
where $\mathcal{N}(i)$ is the neighborhood of the $i_{\text{th}}$ node determined by a cutoff radius and a maximum number of neighbors within that cutoff radius.
The messages $\textbf{m}^{tn}_{ij}$ and attention vectors $\textbf{a}^{tn}_{ij}$ are calculated by a fully connected network from a concatenation of the previous node and edge embeddings. We run $n$ networks for messages and attention coefficients in parallel, each representing one so-called attention head. Here, $\text{FCNNN}^{t,n}_a$ is the network of the $n_{\text{th}}$ attention head at timestep $t$. 
\begin{gather}
    \mathbf{\mathbf{s}}^{t,n}_{ij}= \text{FCNNN}^{t,n}_a(\mathbf{h}^t_i||\mathbf{h}^t_j||\mathbf{e}_{ij})\\
    \mathbf{a}^{t,n}_{ij} =\frac{\exp(s^{t,n}_{ij})}{\sum_j \exp(s^{t,n}_{ij})} \\
    \mathbf{m}^{t,n}_{ij}=\text{FCNN}^{t,n}_m(\mathbf{h}^t_i||\mathbf{h}^t_j||\mathbf{e}_{ij}).\\
\label{eq:combine}
    \mathbf{h}^{t+1}_i = \mathbf{h}^{t}_i +\text{HFCNN}^t_{\theta^t_g}\left(\underset{n}{||}\sum_j\mathbf{a}^n_{ij} \mathbf{m}^n_{ij}\right). 
\end{gather}
In \cref{eq:combine} we combine the messages weighted by the attention coefficients 
through a sum and then average the attention heads. The resulting vector enters a hyper-network that is calculated from the difference of the starting node representation and the node representation at timestep $t$. 
The edges are updated in a similar manner:
\begin{gather}
\mathbf{s}^{e,n}_{ij}= \text{FCNNN}^n_a(\mathbf{h}^t_i||\mathbf{h}^t_j||\mathbf{e}^t_{ij})\\
\mathbf{a}^{e,n}_{ij} =\frac{\exp(\mathbf{s}_{ij})}{\sum_n \exp(\textbf{s}^n_{i})}\\
\mathbf{m}^{e,n}_{ij}= \text{FCNNN}^n(\mathbf{h}^t_i||\mathbf{h}^t_j||\mathbf{e}^t_{ij})\\
\mathbf{e}^{t+1}_{ij} = \mathbf{e}^{t}_{ij}+ \text{FCNN}^{n,t}_{\theta^t_g} \left( \underset{n}{||}\mathbf{a}^{e,n}_{ij} \mathbf{m}^{e,n}_{ij}. \right)\end{gather}
After the last message passing step, the atomic representations are concatenated with a global context vector calculated with a ROOST\cite{goodall2019predicting} model and then combined through another attention layer. Finally, the target quantity is calculated with a residual neural network. 

The networks CGAT-2 and CGAT-3 trained for this publication both used the same hyperparameters, specifically 
  optimizer: AdamW;
  learningrate:  0.000125;
  starting embedding: matscholar-embedding;
  nbr-embedding-size: 512;
  msg-heads: 6;
  batch-size: 512;
  max-nbr: 24;
  epochs: 390;
  loss: L1-loss;
  momentum: 0.9;
  weight-decay: 1e-06;
  atom-fea-len: 128;
  message passing steps: 5;
  roost message passing steps: 3;
  other roost parameters: default;
  vector-attention: True;
  edges: updated;
  learning rate: cyclical;
  learning rate schedule: (0.1, 0.05);
  learning rate period: 130;
  hypernetwork: 3 hidden layers, size 128;
  hypernetwork activ. funct.: $\tanh$;
  FCNN: 1 hidden layer, size 512;
  FCNN activ. funct.: leaky RELU~\cite{LIEW2016718}.

Due to the size of DCGAT-3 we decided to only use validation and test set sizes of 5\% which still encompassed 156483 materials.
The training of each network cost approximately 1000 hours on NVIDIA V100 GPUs. 

\subsection{Transfer Learning}
The high-throughput searches with transfer learning were started with the CGAT-1 network and we performed one round of predictions for the selected crystal structure prototypes. Using a cutoff of 200~\meVatom\ we performed validation calculations with DFT and used the resulting data to transfer learn a separate network for each prototype. Here we reduced the learning rate by a factor of 10 in comparison to the normal training and optimized the network until the validation error converged.

Depending on the number of stable compounds that were found during the next cycle of predictions and the error of the network we performed up to three cycles of transfer learning. As can be seen in Table I in the SI, for all except two prototypes, the MAE was already sufficiently small after one round of transfer learning. Only the garnets and the Ruddlesden–Popper layered perovskites required a second round of data accumulation and training to reach such small MAE.

\section{Data availability statement}
All relevant data used in or resulting from this work is available at Materials Cloud (\url{https://doi.org/10.24435/materialscloud:m7-50}).

\section{Author  Contributions}
JS and NH performed the training of the machines and the machine learning predictions of the distance to the hull; MALM and HCW performed the DFT high-throughput calculations; PJMAC performed the training and the machine learning predictions for the material properties; PB and TFTC performed the calculations of the material properties; MALM and SB directed the research; all authors contributed to the analysis of the results and to the writing of the manuscript.

\section{Competing  Interests}
The authors declare that they have no competing interests.

\section{Acknowledgements}
The authors gratefully acknowledge the Gauss Centre for Supercomputing e.V.
(www.gauss-centre.eu) for funding this project by providing computing time on
the GCS Supercomputer SUPERMUC-NG at Leibniz Supercomputing Centre
(www.lrz.de) under the project pn25co.
TFTC, PJMAC and PB acknowledge financial suport from FCT - Fundação para a Ciência e Tecnologia, Portugal (projects UIDB/04564/2020 and UIDP/04564/2020 and contract 2020.04225.CEECIND) and computational resources provided by the Laboratory for Advanced Computing at University of Coimbra.

\bibliography{bibliography.bib}

\begin{thebibliography}{65}%
\makeatletter
\providecommand \@ifxundefined [1]{%
 \@ifx{#1\undefined}
}%
\providecommand \@ifnum [1]{%
 \ifnum #1\expandafter \@firstoftwo
 \else \expandafter \@secondoftwo
 \fi
}%
\providecommand \@ifx [1]{%
 \ifx #1\expandafter \@firstoftwo
 \else \expandafter \@secondoftwo
 \fi
}%
\providecommand \natexlab [1]{#1}%
\providecommand \enquote  [1]{``#1''}%
\providecommand \bibnamefont  [1]{#1}%
\providecommand \bibfnamefont [1]{#1}%
\providecommand \citenamefont [1]{#1}%
\providecommand \href@noop [0]{\@secondoftwo}%
\providecommand \href [0]{\begingroup \@sanitize@url \@href}%
\providecommand \@href[1]{\@@startlink{#1}\@@href}%
\providecommand \@@href[1]{\endgroup#1\@@endlink}%
\providecommand \@sanitize@url [0]{\catcode `\\12\catcode `\$12\catcode
  `\&12\catcode `\#12\catcode `\^12\catcode `\_12\catcode `\%12\relax}%
\providecommand \@@startlink[1]{}%
\providecommand \@@endlink[0]{}%
\providecommand \url  [0]{\begingroup\@sanitize@url \@url }%
\providecommand \@url [1]{\endgroup\@href {#1}{\urlprefix }}%
\providecommand \urlprefix  [0]{URL }%
\providecommand \Eprint [0]{\href }%
\providecommand \doibase [0]{https://doi.org/}%
\providecommand \selectlanguage [0]{\@gobble}%
\providecommand \bibinfo  [0]{\@secondoftwo}%
\providecommand \bibfield  [0]{\@secondoftwo}%
\providecommand \translation [1]{[#1]}%
\providecommand \BibitemOpen [0]{}%
\providecommand \bibitemStop [0]{}%
\providecommand \bibitemNoStop [0]{.\EOS\space}%
\providecommand \EOS [0]{\spacefactor3000\relax}%
\providecommand \BibitemShut  [1]{\csname bibitem#1\endcsname}%
\let\auto@bib@innerbib\@empty
\bibitem [{\citenamefont {Curtarolo}\ \emph {et~al.}(2012)\citenamefont
  {Curtarolo}, \citenamefont {Setyawan}, \citenamefont {Hart}, \citenamefont
  {Jahnatek}, \citenamefont {Chepulskii}, \citenamefont {Taylor}, \citenamefont
  {Wang}, \citenamefont {Xue}, \citenamefont {Yang}, \citenamefont {Levy},
  \citenamefont {Mehl}, \citenamefont {Stokes}, \citenamefont {Demchenko},\
  and\ \citenamefont {Morgan}}]{aflowlib}%
  \BibitemOpen
  \bibfield  {author} {\bibinfo {author} {\bibfnamefont {S.}~\bibnamefont
  {Curtarolo}}, \bibinfo {author} {\bibfnamefont {W.}~\bibnamefont {Setyawan}},
  \bibinfo {author} {\bibfnamefont {G.~L.}\ \bibnamefont {Hart}}, \bibinfo
  {author} {\bibfnamefont {M.}~\bibnamefont {Jahnatek}}, \bibinfo {author}
  {\bibfnamefont {R.~V.}\ \bibnamefont {Chepulskii}}, \bibinfo {author}
  {\bibfnamefont {R.~H.}\ \bibnamefont {Taylor}}, \bibinfo {author}
  {\bibfnamefont {S.}~\bibnamefont {Wang}}, \bibinfo {author} {\bibfnamefont
  {J.}~\bibnamefont {Xue}}, \bibinfo {author} {\bibfnamefont {K.}~\bibnamefont
  {Yang}}, \bibinfo {author} {\bibfnamefont {O.}~\bibnamefont {Levy}}, \bibinfo
  {author} {\bibfnamefont {M.~J.}\ \bibnamefont {Mehl}}, \bibinfo {author}
  {\bibfnamefont {H.~T.}\ \bibnamefont {Stokes}}, \bibinfo {author}
  {\bibfnamefont {D.~O.}\ \bibnamefont {Demchenko}},\ and\ \bibinfo {author}
  {\bibfnamefont {D.}~\bibnamefont {Morgan}},\ }\bibfield  {title} {\bibinfo
  {title} {{AFLOW}: An automatic framework for high-throughput materials
  discovery},\ }\href
  {https://doi.org/https://doi.org/10.1016/j.commatsci.2012.02.005} {\bibfield
  {journal} {\bibinfo  {journal} {Comput. Mater. Sci.}\ }\textbf {\bibinfo
  {volume} {58}},\ \bibinfo {pages} {218 } (\bibinfo {year}
  {2012})}\BibitemShut {NoStop}%
\bibitem [{\citenamefont {Schmidt}\ \emph {et~al.}(2017)\citenamefont
  {Schmidt}, \citenamefont {Shi}, \citenamefont {Borlido}, \citenamefont
  {Chen}, \citenamefont {Botti},\ and\ \citenamefont {Marques}}]{schmidt2017}%
  \BibitemOpen
  \bibfield  {author} {\bibinfo {author} {\bibfnamefont {J.}~\bibnamefont
  {Schmidt}}, \bibinfo {author} {\bibfnamefont {J.}~\bibnamefont {Shi}},
  \bibinfo {author} {\bibfnamefont {P.}~\bibnamefont {Borlido}}, \bibinfo
  {author} {\bibfnamefont {L.}~\bibnamefont {Chen}}, \bibinfo {author}
  {\bibfnamefont {S.}~\bibnamefont {Botti}},\ and\ \bibinfo {author}
  {\bibfnamefont {M.~A.~L.}\ \bibnamefont {Marques}},\ }\bibfield  {title}
  {\bibinfo {title} {Predicting the thermodynamic stability of solids combining
  density functional theory and machine learning},\ }\href
  {https://doi.org/10.1021/acs.chemmater.7b00156} {\bibfield  {journal}
  {\bibinfo  {journal} {Chem. Mater.}\ }\textbf {\bibinfo {volume} {29}},\
  \bibinfo {pages} {5090} (\bibinfo {year} {2017})}\BibitemShut {NoStop}%
\bibitem [{\citenamefont {Draxl}\ and\ \citenamefont
  {Scheffler}(2018)}]{draxl2018nomad}%
  \BibitemOpen
  \bibfield  {author} {\bibinfo {author} {\bibfnamefont {C.}~\bibnamefont
  {Draxl}}\ and\ \bibinfo {author} {\bibfnamefont {M.}~\bibnamefont
  {Scheffler}},\ }\bibfield  {title} {\bibinfo {title} {{NOMAD}: The {FAIR}
  concept for big data-driven materials science},\ }\href
  {https://doi.org/10.1557/mrs.2018.208} {\bibfield  {journal} {\bibinfo
  {journal} {{MRS} Bull.}\ }\textbf {\bibinfo {volume} {43}},\ \bibinfo {pages}
  {676} (\bibinfo {year} {2018})}\BibitemShut {NoStop}%
\bibitem [{\citenamefont {Jain}\ \emph {et~al.}(2013)\citenamefont {Jain},
  \citenamefont {Ong}, \citenamefont {Hautier}, \citenamefont {Chen},
  \citenamefont {Richards}, \citenamefont {Dacek}, \citenamefont {Cholia},
  \citenamefont {Gunter}, \citenamefont {Skinner}, \citenamefont {Ceder},\ and\
  \citenamefont {Persson}}]{materialsproject}%
  \BibitemOpen
  \bibfield  {author} {\bibinfo {author} {\bibfnamefont {A.}~\bibnamefont
  {Jain}}, \bibinfo {author} {\bibfnamefont {S.~P.}\ \bibnamefont {Ong}},
  \bibinfo {author} {\bibfnamefont {G.}~\bibnamefont {Hautier}}, \bibinfo
  {author} {\bibfnamefont {W.}~\bibnamefont {Chen}}, \bibinfo {author}
  {\bibfnamefont {W.~D.}\ \bibnamefont {Richards}}, \bibinfo {author}
  {\bibfnamefont {S.}~\bibnamefont {Dacek}}, \bibinfo {author} {\bibfnamefont
  {S.}~\bibnamefont {Cholia}}, \bibinfo {author} {\bibfnamefont
  {D.}~\bibnamefont {Gunter}}, \bibinfo {author} {\bibfnamefont
  {D.}~\bibnamefont {Skinner}}, \bibinfo {author} {\bibfnamefont
  {G.}~\bibnamefont {Ceder}},\ and\ \bibinfo {author} {\bibfnamefont {K.~A.}\
  \bibnamefont {Persson}},\ }\bibfield  {title} {\bibinfo {title} {Commentary:
  The materials project: A materials genome approach to accelerating materials
  innovation},\ }\href {https://doi.org/10.1063/1.4812323} {\bibfield
  {journal} {\bibinfo  {journal} {APL Mater.}\ }\textbf {\bibinfo {volume}
  {1}},\ \bibinfo {pages} {011002} (\bibinfo {year} {2013})}\BibitemShut
  {NoStop}%
\bibitem [{\citenamefont {Schmidt}\ \emph {et~al.}(2019)\citenamefont
  {Schmidt}, \citenamefont {Marques}, \citenamefont {Botti},\ and\
  \citenamefont {Marques}}]{ourreview}%
  \BibitemOpen
  \bibfield  {author} {\bibinfo {author} {\bibfnamefont {J.}~\bibnamefont
  {Schmidt}}, \bibinfo {author} {\bibfnamefont {M.~R.~G.}\ \bibnamefont
  {Marques}}, \bibinfo {author} {\bibfnamefont {S.}~\bibnamefont {Botti}},\
  and\ \bibinfo {author} {\bibfnamefont {M.~A.~L.}\ \bibnamefont {Marques}},\
  }\bibfield  {title} {\bibinfo {title} {Recent advances and applications of
  machine learning in solid-state materials science},\ }\href
  {https://doi.org/10.1038/s41524-019-0221-0} {\bibfield  {journal} {\bibinfo
  {journal} {Npj Comput. Mater.}\ }\textbf {\bibinfo {volume} {5}},\ \bibinfo
  {pages} {83} (\bibinfo {year} {2019})}\BibitemShut {NoStop}%
\bibitem [{\citenamefont {Kulik}\ \emph {et~al.}(2022)\citenamefont {Kulik},
  \citenamefont {Hammerschmidt}, \citenamefont {Schmidt}, \citenamefont
  {Botti}, \citenamefont {Marques}, \citenamefont {Boley}, \citenamefont
  {Scheffler}, \citenamefont {Todorovi{\'{c}}}, \citenamefont {Rinke},
  \citenamefont {Oses}, \citenamefont {Smolyanyuk}, \citenamefont {Curtarolo},
  \citenamefont {Tkatchenko}, \citenamefont {Bart{\'{o}}k}, \citenamefont
  {Manzhos}, \citenamefont {Ihara}, \citenamefont {Carrington}, \citenamefont
  {Behler}, \citenamefont {Isayev}, \citenamefont {Veit}, \citenamefont
  {Grisafi}, \citenamefont {Nigam}, \citenamefont {Ceriotti}, \citenamefont
  {Schütt}, \citenamefont {Westermayr}, \citenamefont {Gastegger},
  \citenamefont {Maurer}, \citenamefont {Kalita}, \citenamefont {Burke},
  \citenamefont {Nagai}, \citenamefont {Akashi}, \citenamefont {Sugino},
  \citenamefont {Hermann}, \citenamefont {No{\'{e}}}, \citenamefont {Pilati},
  \citenamefont {Draxl}, \citenamefont {Kuban}, \citenamefont {Rigamonti},
  \citenamefont {Scheidgen}, \citenamefont {Esters}, \citenamefont {Hicks},
  \citenamefont {Toher}, \citenamefont {Balachandran}, \citenamefont {Tamblyn},
  \citenamefont {Whitelam}, \citenamefont {Bellinger},\ and\ \citenamefont
  {Ghiringhelli}}]{roadmap}%
  \BibitemOpen
  \bibfield  {author} {\bibinfo {author} {\bibfnamefont {H.~J.}\ \bibnamefont
  {Kulik}}, \bibinfo {author} {\bibfnamefont {T.}~\bibnamefont
  {Hammerschmidt}}, \bibinfo {author} {\bibfnamefont {J.}~\bibnamefont
  {Schmidt}}, \bibinfo {author} {\bibfnamefont {S.}~\bibnamefont {Botti}},
  \bibinfo {author} {\bibfnamefont {M.~A.~L.}\ \bibnamefont {Marques}},
  \bibinfo {author} {\bibfnamefont {M.}~\bibnamefont {Boley}}, \bibinfo
  {author} {\bibfnamefont {M.}~\bibnamefont {Scheffler}}, \bibinfo {author}
  {\bibfnamefont {M.}~\bibnamefont {Todorovi{\'{c}}}}, \bibinfo {author}
  {\bibfnamefont {P.}~\bibnamefont {Rinke}}, \bibinfo {author} {\bibfnamefont
  {C.}~\bibnamefont {Oses}}, \bibinfo {author} {\bibfnamefont {A.}~\bibnamefont
  {Smolyanyuk}}, \bibinfo {author} {\bibfnamefont {S.}~\bibnamefont
  {Curtarolo}}, \bibinfo {author} {\bibfnamefont {A.}~\bibnamefont
  {Tkatchenko}}, \bibinfo {author} {\bibfnamefont {A.~P.}\ \bibnamefont
  {Bart{\'{o}}k}}, \bibinfo {author} {\bibfnamefont {S.}~\bibnamefont
  {Manzhos}}, \bibinfo {author} {\bibfnamefont {M.}~\bibnamefont {Ihara}},
  \bibinfo {author} {\bibfnamefont {T.}~\bibnamefont {Carrington}}, \bibinfo
  {author} {\bibfnamefont {J.}~\bibnamefont {Behler}}, \bibinfo {author}
  {\bibfnamefont {O.}~\bibnamefont {Isayev}}, \bibinfo {author} {\bibfnamefont
  {M.}~\bibnamefont {Veit}}, \bibinfo {author} {\bibfnamefont {A.}~\bibnamefont
  {Grisafi}}, \bibinfo {author} {\bibfnamefont {J.}~\bibnamefont {Nigam}},
  \bibinfo {author} {\bibfnamefont {M.}~\bibnamefont {Ceriotti}}, \bibinfo
  {author} {\bibfnamefont {K.~T.}\ \bibnamefont {Schütt}}, \bibinfo {author}
  {\bibfnamefont {J.}~\bibnamefont {Westermayr}}, \bibinfo {author}
  {\bibfnamefont {M.}~\bibnamefont {Gastegger}}, \bibinfo {author}
  {\bibfnamefont {R.~J.}\ \bibnamefont {Maurer}}, \bibinfo {author}
  {\bibfnamefont {B.}~\bibnamefont {Kalita}}, \bibinfo {author} {\bibfnamefont
  {K.}~\bibnamefont {Burke}}, \bibinfo {author} {\bibfnamefont
  {R.}~\bibnamefont {Nagai}}, \bibinfo {author} {\bibfnamefont
  {R.}~\bibnamefont {Akashi}}, \bibinfo {author} {\bibfnamefont
  {O.}~\bibnamefont {Sugino}}, \bibinfo {author} {\bibfnamefont
  {J.}~\bibnamefont {Hermann}}, \bibinfo {author} {\bibfnamefont
  {F.}~\bibnamefont {No{\'{e}}}}, \bibinfo {author} {\bibfnamefont
  {S.}~\bibnamefont {Pilati}}, \bibinfo {author} {\bibfnamefont
  {C.}~\bibnamefont {Draxl}}, \bibinfo {author} {\bibfnamefont
  {M.}~\bibnamefont {Kuban}}, \bibinfo {author} {\bibfnamefont
  {S.}~\bibnamefont {Rigamonti}}, \bibinfo {author} {\bibfnamefont
  {M.}~\bibnamefont {Scheidgen}}, \bibinfo {author} {\bibfnamefont
  {M.}~\bibnamefont {Esters}}, \bibinfo {author} {\bibfnamefont
  {D.}~\bibnamefont {Hicks}}, \bibinfo {author} {\bibfnamefont
  {C.}~\bibnamefont {Toher}}, \bibinfo {author} {\bibfnamefont {P.~V.}\
  \bibnamefont {Balachandran}}, \bibinfo {author} {\bibfnamefont
  {I.}~\bibnamefont {Tamblyn}}, \bibinfo {author} {\bibfnamefont
  {S.}~\bibnamefont {Whitelam}}, \bibinfo {author} {\bibfnamefont
  {C.}~\bibnamefont {Bellinger}},\ and\ \bibinfo {author} {\bibfnamefont
  {L.~M.}\ \bibnamefont {Ghiringhelli}},\ }\bibfield  {title} {\bibinfo {title}
  {Roadmap on machine learning in electronic structure},\ }\href
  {https://doi.org/10.1088/2516-1075/ac572f} {\bibfield  {journal} {\bibinfo
  {journal} {Electron. Struct.}\ }\textbf {\bibinfo {volume} {4}},\ \bibinfo
  {pages} {023004} (\bibinfo {year} {2022})}\BibitemShut {NoStop}%
\bibitem [{\citenamefont {Rodrigues}\ \emph {et~al.}(2021)\citenamefont
  {Rodrigues}, \citenamefont {Florea}, \citenamefont {de~Oliveira},
  \citenamefont {Diamond},\ and\ \citenamefont {Oliveira}}]{Rodrigues2021}%
  \BibitemOpen
  \bibfield  {author} {\bibinfo {author} {\bibfnamefont {J.~F.}\ \bibnamefont
  {Rodrigues}}, \bibinfo {author} {\bibfnamefont {L.}~\bibnamefont {Florea}},
  \bibinfo {author} {\bibfnamefont {M.~C.~F.}\ \bibnamefont {de~Oliveira}},
  \bibinfo {author} {\bibfnamefont {D.}~\bibnamefont {Diamond}},\ and\ \bibinfo
  {author} {\bibfnamefont {O.~N.}\ \bibnamefont {Oliveira}},\ }\bibfield
  {title} {\bibinfo {title} {Big data and machine learning for materials
  science},\ }\href {https://doi.org/10.1007/s43939-021-00012-0} {\bibfield
  {journal} {\bibinfo  {journal} {Discov. Mater.}\ }\textbf {\bibinfo {volume}
  {1}},\ \bibinfo {pages} {12} (\bibinfo {year} {2021})}\BibitemShut {NoStop}%
\bibitem [{\citenamefont {Schmidt}\ \emph {et~al.}(2018)\citenamefont
  {Schmidt}, \citenamefont {Chen}, \citenamefont {Botti},\ and\ \citenamefont
  {Marques}}]{jonathan2018}%
  \BibitemOpen
  \bibfield  {author} {\bibinfo {author} {\bibfnamefont {J.}~\bibnamefont
  {Schmidt}}, \bibinfo {author} {\bibfnamefont {L.}~\bibnamefont {Chen}},
  \bibinfo {author} {\bibfnamefont {S.}~\bibnamefont {Botti}},\ and\ \bibinfo
  {author} {\bibfnamefont {M.~A.~L.}\ \bibnamefont {Marques}},\ }\bibfield
  {title} {\bibinfo {title} {Predicting the stability of ternary intermetallics
  with density functional theory and machine learning},\ }\href
  {https://doi.org/10.1063/1.5020223} {\bibfield  {journal} {\bibinfo
  {journal} {J. Chem. Phys.}\ }\textbf {\bibinfo {volume} {148}},\ \bibinfo
  {pages} {241728} (\bibinfo {year} {2018})}\BibitemShut {NoStop}%
\bibitem [{\citenamefont {Ward}\ \emph {et~al.}(2017)\citenamefont {Ward},
  \citenamefont {Liu}, \citenamefont {Krishna}, \citenamefont {Hegde},
  \citenamefont {Agrawal}, \citenamefont {Choudhary},\ and\ \citenamefont
  {Wolverton}}]{60Voronoitessellations}%
  \BibitemOpen
  \bibfield  {author} {\bibinfo {author} {\bibfnamefont {L.}~\bibnamefont
  {Ward}}, \bibinfo {author} {\bibfnamefont {R.}~\bibnamefont {Liu}}, \bibinfo
  {author} {\bibfnamefont {A.}~\bibnamefont {Krishna}}, \bibinfo {author}
  {\bibfnamefont {V.~I.}\ \bibnamefont {Hegde}}, \bibinfo {author}
  {\bibfnamefont {A.}~\bibnamefont {Agrawal}}, \bibinfo {author} {\bibfnamefont
  {A.}~\bibnamefont {Choudhary}},\ and\ \bibinfo {author} {\bibfnamefont
  {C.}~\bibnamefont {Wolverton}},\ }\bibfield  {title} {\bibinfo {title}
  {Including crystal structure attributes in machine learning models of
  formation energies via {Voronoi} tessellations},\ }\href
  {https://doi.org/10.1103/physrevb.96.024104} {\bibfield  {journal} {\bibinfo
  {journal} {Phys. Rev. B}\ }\textbf {\bibinfo {volume} {96}},\ \bibinfo
  {pages} {024104} (\bibinfo {year} {2017})}\BibitemShut {NoStop}%
\bibitem [{\citenamefont {Jha}\ \emph {et~al.}(2018)\citenamefont {Jha},
  \citenamefont {Ward}, \citenamefont {Paul}, \citenamefont {Liao},
  \citenamefont {Choudhary}, \citenamefont {Wolverton},\ and\ \citenamefont
  {Agrawal}}]{ElemNet}%
  \BibitemOpen
  \bibfield  {author} {\bibinfo {author} {\bibfnamefont {D.}~\bibnamefont
  {Jha}}, \bibinfo {author} {\bibfnamefont {L.}~\bibnamefont {Ward}}, \bibinfo
  {author} {\bibfnamefont {A.}~\bibnamefont {Paul}}, \bibinfo {author}
  {\bibfnamefont {W.-k.}\ \bibnamefont {Liao}}, \bibinfo {author}
  {\bibfnamefont {A.}~\bibnamefont {Choudhary}}, \bibinfo {author}
  {\bibfnamefont {C.}~\bibnamefont {Wolverton}},\ and\ \bibinfo {author}
  {\bibfnamefont {A.}~\bibnamefont {Agrawal}},\ }\bibfield  {title} {\bibinfo
  {title} {Elemnet: Deep learning the chemistry of materials from only
  elemental composition},\ }\href {https://doi.org/10.1038/s41598-018-35934-y}
  {\bibfield  {journal} {\bibinfo  {journal} {Sci. Rep.}\ }\textbf {\bibinfo
  {volume} {8}},\ \bibinfo {pages} {17593} (\bibinfo {year}
  {2018})}\BibitemShut {NoStop}%
\bibitem [{\citenamefont {Goodall}\ and\ \citenamefont
  {Lee}(2020)}]{goodall2019predicting}%
  \BibitemOpen
  \bibfield  {author} {\bibinfo {author} {\bibfnamefont {R.~E.~A.}\
  \bibnamefont {Goodall}}\ and\ \bibinfo {author} {\bibfnamefont {A.~A.}\
  \bibnamefont {Lee}},\ }\bibfield  {title} {\bibinfo {title} {Predicting
  materials properties without crystal structure: deep representation learning
  from stoichiometry},\ }\href {https://doi.org/10.1038/s41467-020-19964-7}
  {\bibfield  {journal} {\bibinfo  {journal} {Nat. Commun.}\ }\textbf {\bibinfo
  {volume} {11}},\ \bibinfo {pages} {6280} (\bibinfo {year}
  {2020})}\BibitemShut {NoStop}%
\bibitem [{\citenamefont {Zheng}\ \emph {et~al.}(2018)\citenamefont {Zheng},
  \citenamefont {Zheng},\ and\ \citenamefont {Zhang}}]{CNN2D}%
  \BibitemOpen
  \bibfield  {author} {\bibinfo {author} {\bibfnamefont {X.}~\bibnamefont
  {Zheng}}, \bibinfo {author} {\bibfnamefont {P.}~\bibnamefont {Zheng}},\ and\
  \bibinfo {author} {\bibfnamefont {R.-Z.}\ \bibnamefont {Zhang}},\ }\bibfield
  {title} {\bibinfo {title} {Machine learning material properties from the
  periodic table using convolutional neural networks},\ }\href
  {https://doi.org/10.1039/C8SC02648C} {\bibfield  {journal} {\bibinfo
  {journal} {Chem. Sci.}\ }\textbf {\bibinfo {volume} {9}},\ \bibinfo {pages}
  {8426} (\bibinfo {year} {2018})}\BibitemShut {NoStop}%
\bibitem [{\citenamefont {Zheng}\ \emph {et~al.}(2020)\citenamefont {Zheng},
  \citenamefont {Zheng}, \citenamefont {Zheng}, \citenamefont {Zhang},\ and\
  \citenamefont {Zhang}}]{CNN3Dinput}%
  \BibitemOpen
  \bibfield  {author} {\bibinfo {author} {\bibfnamefont {X.}~\bibnamefont
  {Zheng}}, \bibinfo {author} {\bibfnamefont {P.}~\bibnamefont {Zheng}},
  \bibinfo {author} {\bibfnamefont {L.}~\bibnamefont {Zheng}}, \bibinfo
  {author} {\bibfnamefont {Y.}~\bibnamefont {Zhang}},\ and\ \bibinfo {author}
  {\bibfnamefont {R.-Z.}\ \bibnamefont {Zhang}},\ }\bibfield  {title} {\bibinfo
  {title} {Multi-channel convolutional neural networks for materials properties
  prediction},\ }\href
  {https://doi.org/https://doi.org/10.1016/j.commatsci.2019.109436} {\bibfield
  {journal} {\bibinfo  {journal} {Comput. Mater. Sci.}\ }\textbf {\bibinfo
  {volume} {173}},\ \bibinfo {pages} {109436} (\bibinfo {year}
  {2020})}\BibitemShut {NoStop}%
\bibitem [{\citenamefont {Dunn}\ \emph {et~al.}(2020)\citenamefont {Dunn},
  \citenamefont {Wang}, \citenamefont {Ganose}, \citenamefont {Dopp},\ and\
  \citenamefont {Jain}}]{dunn2020benchmarking}%
  \BibitemOpen
  \bibfield  {author} {\bibinfo {author} {\bibfnamefont {A.}~\bibnamefont
  {Dunn}}, \bibinfo {author} {\bibfnamefont {Q.}~\bibnamefont {Wang}}, \bibinfo
  {author} {\bibfnamefont {A.}~\bibnamefont {Ganose}}, \bibinfo {author}
  {\bibfnamefont {D.}~\bibnamefont {Dopp}},\ and\ \bibinfo {author}
  {\bibfnamefont {A.}~\bibnamefont {Jain}},\ }\bibfield  {title} {\bibinfo
  {title} {Benchmarking materials property prediction methods: the matbench
  test set and automatminer reference algorithm},\ }\href
  {https://doi.org/10.1038/s41524-020-00406-3} {\bibfield  {journal} {\bibinfo
  {journal} {Npj Comput. Mater.}\ }\textbf {\bibinfo {volume} {6}},\ \bibinfo
  {pages} {138} (\bibinfo {year} {2020})}\BibitemShut {NoStop}%
\bibitem [{\citenamefont {Bartel}\ \emph {et~al.}(2020)\citenamefont {Bartel},
  \citenamefont {Trewartha}, \citenamefont {Wang}, \citenamefont {Dunn},
  \citenamefont {Jain},\ and\ \citenamefont {Ceder}}]{bartel2020critical}%
  \BibitemOpen
  \bibfield  {author} {\bibinfo {author} {\bibfnamefont {C.~J.}\ \bibnamefont
  {Bartel}}, \bibinfo {author} {\bibfnamefont {A.}~\bibnamefont {Trewartha}},
  \bibinfo {author} {\bibfnamefont {Q.}~\bibnamefont {Wang}}, \bibinfo {author}
  {\bibfnamefont {A.}~\bibnamefont {Dunn}}, \bibinfo {author} {\bibfnamefont
  {A.}~\bibnamefont {Jain}},\ and\ \bibinfo {author} {\bibfnamefont
  {G.}~\bibnamefont {Ceder}},\ }\bibfield  {title} {\bibinfo {title} {A
  critical examination of compound stability predictions from machine-learned
  formation energies},\ }\href {https://doi.org/10.1038/s41524-020-00362-y}
  {\bibfield  {journal} {\bibinfo  {journal} {Npj Comput. Mater.}\ }\textbf
  {\bibinfo {volume} {6}},\ \bibinfo {pages} {97} (\bibinfo {year}
  {2020})}\BibitemShut {NoStop}%
\bibitem [{\citenamefont {Schmidt}\ \emph {et~al.}(2021)\citenamefont
  {Schmidt}, \citenamefont {Pettersson}, \citenamefont {Verdozzi},
  \citenamefont {Botti},\ and\ \citenamefont {Marques}}]{CGAT}%
  \BibitemOpen
  \bibfield  {author} {\bibinfo {author} {\bibfnamefont {J.}~\bibnamefont
  {Schmidt}}, \bibinfo {author} {\bibfnamefont {L.}~\bibnamefont {Pettersson}},
  \bibinfo {author} {\bibfnamefont {C.}~\bibnamefont {Verdozzi}}, \bibinfo
  {author} {\bibfnamefont {S.}~\bibnamefont {Botti}},\ and\ \bibinfo {author}
  {\bibfnamefont {M.~A.~L.}\ \bibnamefont {Marques}},\ }\bibfield  {title}
  {\bibinfo {title} {Crystal graph attention networks for the prediction of
  stable materials},\ }\href {https://doi.org/10.1126/sciadv.abi7948}
  {\bibfield  {journal} {\bibinfo  {journal} {Sci. Adv.}\ }\textbf {\bibinfo
  {volume} {7}},\ \bibinfo {pages} {eabi7948} (\bibinfo {year}
  {2021})}\BibitemShut {NoStop}%
\bibitem [{\citenamefont {Goodall}\ \emph {et~al.}(2021)\citenamefont
  {Goodall}, \citenamefont {Parackal}, \citenamefont {Faber}, \citenamefont
  {Armiento},\ and\ \citenamefont {Lee}}]{goodall2021rapid}%
  \BibitemOpen
  \bibfield  {author} {\bibinfo {author} {\bibfnamefont {R.~E.}\ \bibnamefont
  {Goodall}}, \bibinfo {author} {\bibfnamefont {A.~S.}\ \bibnamefont
  {Parackal}}, \bibinfo {author} {\bibfnamefont {F.~A.}\ \bibnamefont {Faber}},
  \bibinfo {author} {\bibfnamefont {R.}~\bibnamefont {Armiento}},\ and\
  \bibinfo {author} {\bibfnamefont {A.~A.}\ \bibnamefont {Lee}},\ }\href@noop
  {} {\bibinfo {title} {Rapid discovery of novel materials by coordinate-free
  coarse graining}},\ \bibinfo {howpublished} {Preprint at
  https://arxiv.org/abs/2106.11132} (\bibinfo {year} {2021})\BibitemShut
  {NoStop}%
\bibitem [{\citenamefont {Beker}\ \emph {et~al.}(2022)\citenamefont {Beker},
  \citenamefont {Roszak}, \citenamefont {Wołos}, \citenamefont {Angello},
  \citenamefont {Rathore}, \citenamefont {Burke},\ and\ \citenamefont
  {Grzybowski}}]{doi:10.1021/jacs.1c12005}%
  \BibitemOpen
  \bibfield  {author} {\bibinfo {author} {\bibfnamefont {W.}~\bibnamefont
  {Beker}}, \bibinfo {author} {\bibfnamefont {R.}~\bibnamefont {Roszak}},
  \bibinfo {author} {\bibfnamefont {A.}~\bibnamefont {Wołos}}, \bibinfo
  {author} {\bibfnamefont {N.~H.}\ \bibnamefont {Angello}}, \bibinfo {author}
  {\bibfnamefont {V.}~\bibnamefont {Rathore}}, \bibinfo {author} {\bibfnamefont
  {M.~D.}\ \bibnamefont {Burke}},\ and\ \bibinfo {author} {\bibfnamefont
  {B.~A.}\ \bibnamefont {Grzybowski}},\ }\bibfield  {title} {\bibinfo {title}
  {Machine learning may sometimes simply capture literature popularity trends:
  A case study of heterocyclic {S}uzuki–{M}iyaura coupling},\ }\href
  {https://doi.org/10.1021/jacs.1c12005} {\bibfield  {journal} {\bibinfo
  {journal} {J. Am. Chem. Soc.}\ }\textbf {\bibinfo {volume} {144}},\ \bibinfo
  {pages} {4819} (\bibinfo {year} {2022})}\BibitemShut {NoStop}%
\bibitem [{\citenamefont {Restrepo}(2022)}]{D2DD00030J}%
  \BibitemOpen
  \bibfield  {author} {\bibinfo {author} {\bibfnamefont {G.}~\bibnamefont
  {Restrepo}},\ }\bibfield  {title} {\bibinfo {title} {Chemical space:
  limits{,} evolution and modelling of an object bigger than our universal
  library},\ }\href {https://doi.org/10.1039/D2DD00030J} {\bibfield  {journal}
  {\bibinfo  {journal} {Digital Discovery}\ ,\ } (\bibinfo {year}
  {2022})}\BibitemShut {NoStop}%
\bibitem [{\citenamefont {Wang}\ \emph
  {et~al.}(2021{\natexlab{a}})\citenamefont {Wang}, \citenamefont {Schmidt},
  \citenamefont {Botti},\ and\ \citenamefont {Marques}}]{wang2021high}%
  \BibitemOpen
  \bibfield  {author} {\bibinfo {author} {\bibfnamefont {H.}~\bibnamefont
  {Wang}}, \bibinfo {author} {\bibfnamefont {J.}~\bibnamefont {Schmidt}},
  \bibinfo {author} {\bibfnamefont {S.}~\bibnamefont {Botti}},\ and\ \bibinfo
  {author} {\bibfnamefont {M.~A.~L.}\ \bibnamefont {Marques}},\ }\bibfield
  {title} {\bibinfo {title} {A high-throughput study of oxynitride{,}
  oxyfluoride and nitrofluoride perovskites},\ }\href
  {https://doi.org/10.1039/D0TA10781F} {\bibfield  {journal} {\bibinfo
  {journal} {J. Mater. Chem. A}\ }\textbf {\bibinfo {volume} {9}},\ \bibinfo
  {pages} {8501} (\bibinfo {year} {2021}{\natexlab{a}})}\BibitemShut {NoStop}%
\bibitem [{\citenamefont {Oses}\ \emph {et~al.}(2018)\citenamefont {Oses},
  \citenamefont {Gossett}, \citenamefont {Hicks}, \citenamefont {Rose},
  \citenamefont {Mehl}, \citenamefont {Perim}, \citenamefont {Takeuchi},
  \citenamefont {Sanvito}, \citenamefont {Scheffler}, \citenamefont {Lederer},
  \citenamefont {Levy}, \citenamefont {Toher},\ and\ \citenamefont
  {Curtarolo}}]{aflowchull}%
  \BibitemOpen
  \bibfield  {author} {\bibinfo {author} {\bibfnamefont {C.}~\bibnamefont
  {Oses}}, \bibinfo {author} {\bibfnamefont {E.}~\bibnamefont {Gossett}},
  \bibinfo {author} {\bibfnamefont {D.~V.}\ \bibnamefont {Hicks}}, \bibinfo
  {author} {\bibfnamefont {F.}~\bibnamefont {Rose}}, \bibinfo {author}
  {\bibfnamefont {M.~J.}\ \bibnamefont {Mehl}}, \bibinfo {author}
  {\bibfnamefont {E.}~\bibnamefont {Perim}}, \bibinfo {author} {\bibfnamefont
  {I.}~\bibnamefont {Takeuchi}}, \bibinfo {author} {\bibfnamefont
  {S.}~\bibnamefont {Sanvito}}, \bibinfo {author} {\bibfnamefont
  {M.}~\bibnamefont {Scheffler}}, \bibinfo {author} {\bibfnamefont
  {Y.}~\bibnamefont {Lederer}}, \bibinfo {author} {\bibfnamefont
  {O.}~\bibnamefont {Levy}}, \bibinfo {author} {\bibfnamefont {C.}~\bibnamefont
  {Toher}},\ and\ \bibinfo {author} {\bibfnamefont {S.}~\bibnamefont
  {Curtarolo}},\ }\bibfield  {title} {\bibinfo {title} {Aflow-chull:
  Cloud-oriented platform for autonomous phase stability analysis.},\ }\href
  {https://doi.org/10.1021/acs.jcim.8b00393} {\bibfield  {journal} {\bibinfo
  {journal} {J. Chem. Inf. Model.}\ }\textbf {\bibinfo {volume} {58 12}},\
  \bibinfo {pages} {2477} (\bibinfo {year} {2018})}\BibitemShut {NoStop}%
\bibitem [{\citenamefont {Ong}\ \emph {et~al.}(2013)\citenamefont {Ong},
  \citenamefont {Richards}, \citenamefont {Jain}, \citenamefont {Hautier},
  \citenamefont {Kocher}, \citenamefont {Cholia}, \citenamefont {Gunter},
  \citenamefont {Chevrier}, \citenamefont {Persson},\ and\ \citenamefont
  {Ceder}}]{pymatgen2013CompMatSci}%
  \BibitemOpen
  \bibfield  {author} {\bibinfo {author} {\bibfnamefont {S.~P.}\ \bibnamefont
  {Ong}}, \bibinfo {author} {\bibfnamefont {W.~D.}\ \bibnamefont {Richards}},
  \bibinfo {author} {\bibfnamefont {A.}~\bibnamefont {Jain}}, \bibinfo {author}
  {\bibfnamefont {G.}~\bibnamefont {Hautier}}, \bibinfo {author} {\bibfnamefont
  {M.}~\bibnamefont {Kocher}}, \bibinfo {author} {\bibfnamefont
  {S.}~\bibnamefont {Cholia}}, \bibinfo {author} {\bibfnamefont
  {D.}~\bibnamefont {Gunter}}, \bibinfo {author} {\bibfnamefont {V.~L.}\
  \bibnamefont {Chevrier}}, \bibinfo {author} {\bibfnamefont {K.~A.}\
  \bibnamefont {Persson}},\ and\ \bibinfo {author} {\bibfnamefont
  {G.}~\bibnamefont {Ceder}},\ }\bibfield  {title} {\bibinfo {title} {Python
  materials genomics (pymatgen): A robust, open-source python library for
  materials analysis},\ }\href
  {https://doi.org/https://doi.org/10.1016/j.commatsci.2012.10.028} {\bibfield
  {journal} {\bibinfo  {journal} {Comput. Mater. Sci.}\ }\textbf {\bibinfo
  {volume} {68}},\ \bibinfo {pages} {314} (\bibinfo {year} {2013})}\BibitemShut
  {NoStop}%
\bibitem [{\citenamefont {Wang}\ \emph
  {et~al.}(2021{\natexlab{b}})\citenamefont {Wang}, \citenamefont {Botti},\
  and\ \citenamefont {Marques}}]{haichen}%
  \BibitemOpen
  \bibfield  {author} {\bibinfo {author} {\bibfnamefont {H.}~\bibnamefont
  {Wang}}, \bibinfo {author} {\bibfnamefont {S.}~\bibnamefont {Botti}},\ and\
  \bibinfo {author} {\bibfnamefont {M.~A.~L.}\ \bibnamefont {Marques}},\
  }\bibfield  {title} {\bibinfo {title} {Predicting stable crystalline
  compounds using chemical similarity},\ }\href
  {https://doi.org/10.1038/s41524-020-00481-6} {\bibfield  {journal} {\bibinfo
  {journal} {Npj Comput. Mater.}\ }\textbf {\bibinfo {volume} {7}},\ \bibinfo
  {pages} {12} (\bibinfo {year} {2021}{\natexlab{b}})}\BibitemShut {NoStop}%
\bibitem [{\citenamefont {Xie}\ and\ \citenamefont
  {Grossman}(2018)}]{68crystalgraphconvolution}%
  \BibitemOpen
  \bibfield  {author} {\bibinfo {author} {\bibfnamefont {T.}~\bibnamefont
  {Xie}}\ and\ \bibinfo {author} {\bibfnamefont {J.~C.}\ \bibnamefont
  {Grossman}},\ }\bibfield  {title} {\bibinfo {title} {Crystal graph
  convolutional neural networks for an accurate and interpretable prediction of
  material properties},\ }\href
  {https://doi.org/10.1103/physrevlett.120.145301} {\bibfield  {journal}
  {\bibinfo  {journal} {Phys. Rev. Lett.}\ }\textbf {\bibinfo {volume} {120}},\
  \bibinfo {pages} {145301} (\bibinfo {year} {2018})}\BibitemShut {NoStop}%
\bibitem [{\citenamefont {Mouhat}\ and\ \citenamefont
  {Coudert}(2014)}]{PhysRevB.90.224104}%
  \BibitemOpen
  \bibfield  {author} {\bibinfo {author} {\bibfnamefont {F.}~\bibnamefont
  {Mouhat}}\ and\ \bibinfo {author} {\bibfnamefont {F.~m. c.-X.}\ \bibnamefont
  {Coudert}},\ }\bibfield  {title} {\bibinfo {title} {Necessary and sufficient
  elastic stability conditions in various crystal systems},\ }\href
  {https://doi.org/10.1103/PhysRevB.90.224104} {\bibfield  {journal} {\bibinfo
  {journal} {Phys. Rev. B}\ }\textbf {\bibinfo {volume} {90}},\ \bibinfo
  {pages} {224104} (\bibinfo {year} {2014})}\BibitemShut {NoStop}%
\bibitem [{\citenamefont {Born}(1940)}]{born_1940}%
  \BibitemOpen
  \bibfield  {author} {\bibinfo {author} {\bibfnamefont {M.}~\bibnamefont
  {Born}},\ }\bibfield  {title} {\bibinfo {title} {On the stability of crystal
  lattices. {I}},\ }\href {https://doi.org/10.1017/S0305004100017138}
  {\bibfield  {journal} {\bibinfo  {journal} {Math. Proc. Camb. Philos. Soc.}\
  }\textbf {\bibinfo {volume} {36}},\ \bibinfo {pages} {160–172} (\bibinfo
  {year} {1940})}\BibitemShut {NoStop}%
\bibitem [{\citenamefont {Momma}\ and\ \citenamefont {Izumi}(2011)}]{vesta}%
  \BibitemOpen
  \bibfield  {author} {\bibinfo {author} {\bibfnamefont {K.}~\bibnamefont
  {Momma}}\ and\ \bibinfo {author} {\bibfnamefont {F.}~\bibnamefont {Izumi}},\
  }\bibfield  {title} {\bibinfo {title} {{{\it VESTA3} for three-dimensional
  visualization of crystal, volumetric and morphology data}},\ }\href
  {https://doi.org/10.1107/S0021889811038970} {\bibfield  {journal} {\bibinfo
  {journal} {J. Appl. Cryst.}\ }\textbf {\bibinfo {volume} {44}},\ \bibinfo
  {pages} {1272} (\bibinfo {year} {2011})}\BibitemShut {NoStop}%
\bibitem [{\citenamefont {Hill}(1952)}]{0370-1298-65-5-307}%
  \BibitemOpen
  \bibfield  {author} {\bibinfo {author} {\bibfnamefont {R.}~\bibnamefont
  {Hill}},\ }\bibfield  {title} {\bibinfo {title} {The elastic behaviour of a
  crystalline aggregate},\ }\href {https://doi.org/10.1088/0370-1298/65/5/307}
  {\bibfield  {journal} {\bibinfo  {journal} {Proc. Phys. Soc. A}\ }\textbf
  {\bibinfo {volume} {65}},\ \bibinfo {pages} {349} (\bibinfo {year}
  {1952})}\BibitemShut {NoStop}%
\bibitem [{\citenamefont {Smith}\ and\ \citenamefont
  {Sandly}(1922)}]{smith1922ProcIntsMechEng}%
  \BibitemOpen
  \bibfield  {author} {\bibinfo {author} {\bibfnamefont {R.~L.}\ \bibnamefont
  {Smith}}\ and\ \bibinfo {author} {\bibfnamefont {G.~E.}\ \bibnamefont
  {Sandly}},\ }\bibfield  {title} {\bibinfo {title} {An accurate method of
  determining the hardness of metals, with particular reference to those of a
  high degree of hardness},\ }\href
  {https://doi.org/{10.1243/PIME\_PROC\_1922\_102\_033\_02}} {\bibfield
  {journal} {\bibinfo  {journal} {Proc. Inst. Mech. Eng.}\ }\textbf {\bibinfo
  {volume} {102}},\ \bibinfo {pages} {623} (\bibinfo {year}
  {1922})}\BibitemShut {NoStop}%
\bibitem [{\citenamefont {Mazhnik}\ and\ \citenamefont
  {Oganov}(2019)}]{mazhnik2019JAP}%
  \BibitemOpen
  \bibfield  {author} {\bibinfo {author} {\bibfnamefont {E.}~\bibnamefont
  {Mazhnik}}\ and\ \bibinfo {author} {\bibfnamefont {A.~R.}\ \bibnamefont
  {Oganov}},\ }\bibfield  {title} {\bibinfo {title} {A model of hardness and
  fracture toughness of solids},\ }\href {https://doi.org/10.1063/1.5113622}
  {\bibfield  {journal} {\bibinfo  {journal} {J. Appl. Phys.}\ }\textbf
  {\bibinfo {volume} {126}},\ \bibinfo {pages} {125109} (\bibinfo {year}
  {2019})}\BibitemShut {NoStop}%
\bibitem [{\citenamefont {Zuo}\ \emph {et~al.}(2021)\citenamefont {Zuo},
  \citenamefont {Qin}, \citenamefont {Chen}, \citenamefont {Ye}, \citenamefont
  {Li}, \citenamefont {Luo},\ and\ \citenamefont {Ong}}]{zuo2021accelerating}%
  \BibitemOpen
  \bibfield  {author} {\bibinfo {author} {\bibfnamefont {Y.}~\bibnamefont
  {Zuo}}, \bibinfo {author} {\bibfnamefont {M.}~\bibnamefont {Qin}}, \bibinfo
  {author} {\bibfnamefont {C.}~\bibnamefont {Chen}}, \bibinfo {author}
  {\bibfnamefont {W.}~\bibnamefont {Ye}}, \bibinfo {author} {\bibfnamefont
  {X.}~\bibnamefont {Li}}, \bibinfo {author} {\bibfnamefont {J.}~\bibnamefont
  {Luo}},\ and\ \bibinfo {author} {\bibfnamefont {S.~P.}\ \bibnamefont {Ong}},\
  }\bibfield  {title} {\bibinfo {title} {Accelerating materials discovery with
  bayesian optimization and graph deep learning},\ }\href@noop {} {\bibfield
  {journal} {\bibinfo  {journal} {Mater. Today}\ }\textbf {\bibinfo {volume}
  {51}},\ \bibinfo {pages} {126} (\bibinfo {year} {2021})}\BibitemShut
  {NoStop}%
\bibitem [{\citenamefont {Chen}\ \emph {et~al.}(2019)\citenamefont {Chen},
  \citenamefont {Ye}, \citenamefont {Zuo}, \citenamefont {Zheng},\ and\
  \citenamefont {Ong}}]{megnet}%
  \BibitemOpen
  \bibfield  {author} {\bibinfo {author} {\bibfnamefont {C.}~\bibnamefont
  {Chen}}, \bibinfo {author} {\bibfnamefont {W.}~\bibnamefont {Ye}}, \bibinfo
  {author} {\bibfnamefont {Y.}~\bibnamefont {Zuo}}, \bibinfo {author}
  {\bibfnamefont {C.}~\bibnamefont {Zheng}},\ and\ \bibinfo {author}
  {\bibfnamefont {S.~P.}\ \bibnamefont {Ong}},\ }\bibfield  {title} {\bibinfo
  {title} {Graph networks as a universal machine learning framework for
  molecules and crystals},\ }\href
  {https://doi.org/10.1021/acs.chemmater.9b01294} {\bibfield  {journal}
  {\bibinfo  {journal} {Chem. Mater.}\ }\textbf {\bibinfo {volume} {31}},\
  \bibinfo {pages} {3564} (\bibinfo {year} {2019})}\BibitemShut {NoStop}%
\bibitem [{\citenamefont {Akopov}\ \emph {et~al.}(2018)\citenamefont {Akopov},
  \citenamefont {Pangilinan}, \citenamefont {Mohammadi},\ and\ \citenamefont
  {Kaner}}]{ultrahardboridesperspective2018}%
  \BibitemOpen
  \bibfield  {author} {\bibinfo {author} {\bibfnamefont {G.}~\bibnamefont
  {Akopov}}, \bibinfo {author} {\bibfnamefont {L.~E.}\ \bibnamefont
  {Pangilinan}}, \bibinfo {author} {\bibfnamefont {R.}~\bibnamefont
  {Mohammadi}},\ and\ \bibinfo {author} {\bibfnamefont {R.~B.}\ \bibnamefont
  {Kaner}},\ }\bibfield  {title} {\bibinfo {title} {Perspective: Superhard
  metal borides: A look forward},\ }\href {https://doi.org/10.1063/1.5040763}
  {\bibfield  {journal} {\bibinfo  {journal} {APL Mater.}\ }\textbf {\bibinfo
  {volume} {6}},\ \bibinfo {pages} {070901} (\bibinfo {year}
  {2018})}\BibitemShut {NoStop}%
\bibitem [{\citenamefont {Pangilinan}\ \emph {et~al.}(2022)\citenamefont
  {Pangilinan}, \citenamefont {Hu}, \citenamefont {Hamilton}, \citenamefont
  {Tolbert},\ and\ \citenamefont {Kaner}}]{Pangilinan2022}%
  \BibitemOpen
  \bibfield  {author} {\bibinfo {author} {\bibfnamefont {L.~E.}\ \bibnamefont
  {Pangilinan}}, \bibinfo {author} {\bibfnamefont {S.}~\bibnamefont {Hu}},
  \bibinfo {author} {\bibfnamefont {S.~G.}\ \bibnamefont {Hamilton}}, \bibinfo
  {author} {\bibfnamefont {S.~H.}\ \bibnamefont {Tolbert}},\ and\ \bibinfo
  {author} {\bibfnamefont {R.~B.}\ \bibnamefont {Kaner}},\ }\bibfield  {title}
  {\bibinfo {title} {Hardening effects in superhard transition-metal borides},\
  }\href {https://doi.org/10.1021/accountsmr.1c00192} {\bibfield  {journal}
  {\bibinfo  {journal} {Acc. Mater. Res.}\ }\textbf {\bibinfo {volume} {3}},\
  \bibinfo {pages} {100} (\bibinfo {year} {2022})}\BibitemShut {NoStop}%
\bibitem [{\citenamefont {Tawfik}\ \emph {et~al.}(2022)\citenamefont {Tawfik},
  \citenamefont {Nguyen}, \citenamefont {Tran}, \citenamefont {Walsh},\ and\
  \citenamefont {Venkatesh}}]{tawfik2022}%
  \BibitemOpen
  \bibfield  {author} {\bibinfo {author} {\bibfnamefont {S.~A.}\ \bibnamefont
  {Tawfik}}, \bibinfo {author} {\bibfnamefont {P.}~\bibnamefont {Nguyen}},
  \bibinfo {author} {\bibfnamefont {T.}~\bibnamefont {Tran}}, \bibinfo {author}
  {\bibfnamefont {T.~R.}\ \bibnamefont {Walsh}},\ and\ \bibinfo {author}
  {\bibfnamefont {S.}~\bibnamefont {Venkatesh}},\ }\bibfield  {title} {\bibinfo
  {title} {Machine learning-aided exploration of ultrahard materials},\ }\href
  {https://doi.org/10.1021/acs.jpcc.2c03926} {\bibfield  {journal} {\bibinfo
  {journal} {J. Phys. Chem. C}\ }\textbf {\bibinfo {volume} {126}},\ \bibinfo
  {pages} {15952} (\bibinfo {year} {2022})}\BibitemShut {NoStop}%
\bibitem [{\citenamefont {Yeung}\ \emph {et~al.}(2016)\citenamefont {Yeung},
  \citenamefont {Mohammadi},\ and\ \citenamefont
  {Kaner}}]{superhardreview2016}%
  \BibitemOpen
  \bibfield  {author} {\bibinfo {author} {\bibfnamefont {M.~T.}\ \bibnamefont
  {Yeung}}, \bibinfo {author} {\bibfnamefont {R.}~\bibnamefont {Mohammadi}},\
  and\ \bibinfo {author} {\bibfnamefont {R.~B.}\ \bibnamefont {Kaner}},\
  }\bibfield  {title} {\bibinfo {title} {Ultraincompressible, superhard
  materials},\ }\href {https://doi.org/10.1146/annurev-matsci-070115-032148}
  {\bibfield  {journal} {\bibinfo  {journal} {Annu. Rev. Mater. Res.}\ }\textbf
  {\bibinfo {volume} {46}},\ \bibinfo {pages} {465} (\bibinfo {year}
  {2016})}\BibitemShut {NoStop}%
\bibitem [{\citenamefont {Lilia}\ \emph {et~al.}(2022)\citenamefont {Lilia},
  \citenamefont {Hennig}, \citenamefont {Hirschfeld}, \citenamefont {Profeta},
  \citenamefont {Sanna}, \citenamefont {Zurek}, \citenamefont {Pickett},
  \citenamefont {Amsler}, \citenamefont {Dias}, \citenamefont {Eremets},
  \citenamefont {Heil}, \citenamefont {Hemley}, \citenamefont {Liu},
  \citenamefont {Ma}, \citenamefont {Pierleoni}, \citenamefont {Kolmogorov},
  \citenamefont {Rybin}, \citenamefont {Novoselov}, \citenamefont {Anisimov},
  \citenamefont {Oganov}, \citenamefont {Pickard}, \citenamefont {Bi},
  \citenamefont {Arita}, \citenamefont {Errea}, \citenamefont {Pellegrini},
  \citenamefont {Requist}, \citenamefont {Gross}, \citenamefont {Margine},
  \citenamefont {Xie}, \citenamefont {Quan}, \citenamefont {Hire},
  \citenamefont {Fanfarillo}, \citenamefont {Stewart}, \citenamefont {Hamlin},
  \citenamefont {Stanev}, \citenamefont {Gonnelli}, \citenamefont {Piatti},
  \citenamefont {Romanin}, \citenamefont {Daghero},\ and\ \citenamefont
  {Valenti}}]{Lilia_2022_supercond_roadmap}%
  \BibitemOpen
  \bibfield  {author} {\bibinfo {author} {\bibfnamefont {B.}~\bibnamefont
  {Lilia}}, \bibinfo {author} {\bibfnamefont {R.}~\bibnamefont {Hennig}},
  \bibinfo {author} {\bibfnamefont {P.}~\bibnamefont {Hirschfeld}}, \bibinfo
  {author} {\bibfnamefont {G.}~\bibnamefont {Profeta}}, \bibinfo {author}
  {\bibfnamefont {A.}~\bibnamefont {Sanna}}, \bibinfo {author} {\bibfnamefont
  {E.}~\bibnamefont {Zurek}}, \bibinfo {author} {\bibfnamefont {W.~E.}\
  \bibnamefont {Pickett}}, \bibinfo {author} {\bibfnamefont {M.}~\bibnamefont
  {Amsler}}, \bibinfo {author} {\bibfnamefont {R.}~\bibnamefont {Dias}},
  \bibinfo {author} {\bibfnamefont {M.~I.}\ \bibnamefont {Eremets}}, \bibinfo
  {author} {\bibfnamefont {C.}~\bibnamefont {Heil}}, \bibinfo {author}
  {\bibfnamefont {R.~J.}\ \bibnamefont {Hemley}}, \bibinfo {author}
  {\bibfnamefont {H.}~\bibnamefont {Liu}}, \bibinfo {author} {\bibfnamefont
  {Y.}~\bibnamefont {Ma}}, \bibinfo {author} {\bibfnamefont {C.}~\bibnamefont
  {Pierleoni}}, \bibinfo {author} {\bibfnamefont {A.~N.}\ \bibnamefont
  {Kolmogorov}}, \bibinfo {author} {\bibfnamefont {N.}~\bibnamefont {Rybin}},
  \bibinfo {author} {\bibfnamefont {D.}~\bibnamefont {Novoselov}}, \bibinfo
  {author} {\bibfnamefont {V.}~\bibnamefont {Anisimov}}, \bibinfo {author}
  {\bibfnamefont {A.~R.}\ \bibnamefont {Oganov}}, \bibinfo {author}
  {\bibfnamefont {C.~J.}\ \bibnamefont {Pickard}}, \bibinfo {author}
  {\bibfnamefont {T.}~\bibnamefont {Bi}}, \bibinfo {author} {\bibfnamefont
  {R.}~\bibnamefont {Arita}}, \bibinfo {author} {\bibfnamefont
  {I.}~\bibnamefont {Errea}}, \bibinfo {author} {\bibfnamefont
  {C.}~\bibnamefont {Pellegrini}}, \bibinfo {author} {\bibfnamefont
  {R.}~\bibnamefont {Requist}}, \bibinfo {author} {\bibfnamefont {E.~K.~U.}\
  \bibnamefont {Gross}}, \bibinfo {author} {\bibfnamefont {E.~R.}\ \bibnamefont
  {Margine}}, \bibinfo {author} {\bibfnamefont {S.~R.}\ \bibnamefont {Xie}},
  \bibinfo {author} {\bibfnamefont {Y.}~\bibnamefont {Quan}}, \bibinfo {author}
  {\bibfnamefont {A.}~\bibnamefont {Hire}}, \bibinfo {author} {\bibfnamefont
  {L.}~\bibnamefont {Fanfarillo}}, \bibinfo {author} {\bibfnamefont {G.~R.}\
  \bibnamefont {Stewart}}, \bibinfo {author} {\bibfnamefont {J.~J.}\
  \bibnamefont {Hamlin}}, \bibinfo {author} {\bibfnamefont {V.}~\bibnamefont
  {Stanev}}, \bibinfo {author} {\bibfnamefont {R.~S.}\ \bibnamefont
  {Gonnelli}}, \bibinfo {author} {\bibfnamefont {E.}~\bibnamefont {Piatti}},
  \bibinfo {author} {\bibfnamefont {D.}~\bibnamefont {Romanin}}, \bibinfo
  {author} {\bibfnamefont {D.}~\bibnamefont {Daghero}},\ and\ \bibinfo {author}
  {\bibfnamefont {R.}~\bibnamefont {Valenti}},\ }\bibfield  {title} {\bibinfo
  {title} {The 2021 room-temperature superconductivity roadmap},\ }\href
  {https://doi.org/10.1088/1361-648x/ac2864} {\bibfield  {journal} {\bibinfo
  {journal} {J. Phys.: Condens. Matter.}\ }\textbf {\bibinfo {volume} {34}},\
  \bibinfo {pages} {183002} (\bibinfo {year} {2022})}\BibitemShut {NoStop}%
\bibitem [{\citenamefont {McMillan}(1968)}]{mcmillan1968PRB}%
  \BibitemOpen
  \bibfield  {author} {\bibinfo {author} {\bibfnamefont {W.~L.}\ \bibnamefont
  {McMillan}},\ }\bibfield  {title} {\bibinfo {title} {Transition temperature
  of strong-coupled superconductors},\ }\href
  {https://doi.org/10.1103/PhysRev.167.331} {\bibfield  {journal} {\bibinfo
  {journal} {Phys. Rev.}\ }\textbf {\bibinfo {volume} {167}},\ \bibinfo {pages}
  {331} (\bibinfo {year} {1968})}\BibitemShut {NoStop}%
\bibitem [{\citenamefont {Choudhary}\ and\ \citenamefont
  {Garrity}(2022)}]{choudhary2022arxiv}%
  \BibitemOpen
  \bibfield  {author} {\bibinfo {author} {\bibfnamefont {K.}~\bibnamefont
  {Choudhary}}\ and\ \bibinfo {author} {\bibfnamefont {K.}~\bibnamefont
  {Garrity}},\ }\href@noop {} {\bibinfo {title} {Designing high-tc
  superconductors with bcs-inspired screening, density functional theory and
  deep-learning}},\ \bibinfo {howpublished} {Preprint at
  https://arxiv.org/abs/2205.00060} (\bibinfo {year} {2022})\BibitemShut
  {NoStop}%
\bibitem [{\citenamefont {Anderson}(1963)}]{anderson1963JPhysChemSol}%
  \BibitemOpen
  \bibfield  {author} {\bibinfo {author} {\bibfnamefont {O.~L.}\ \bibnamefont
  {Anderson}},\ }\bibfield  {title} {\bibinfo {title} {A simplified method for
  calculating the debye temperature from elastic constants},\ }\href
  {https://doi.org/https://doi.org/10.1016/0022-3697(63)90067-2} {\bibfield
  {journal} {\bibinfo  {journal} {J. Phys. Chem. Solids}\ }\textbf {\bibinfo
  {volume} {24}},\ \bibinfo {pages} {909} (\bibinfo {year} {1963})}\BibitemShut
  {NoStop}%
\bibitem [{\citenamefont {Stein}\ and\ \citenamefont
  {Leineweber}(2021)}]{10.1007/s10853-020-05509-2}%
  \BibitemOpen
  \bibfield  {author} {\bibinfo {author} {\bibfnamefont {F.}~\bibnamefont
  {Stein}}\ and\ \bibinfo {author} {\bibfnamefont {A.}~\bibnamefont
  {Leineweber}},\ }\bibfield  {title} {\bibinfo {title} {Laves phases: a review
  of their functional and structural applications and an improved fundamental
  understanding of stability and properties},\ }\href
  {https://doi.org/10.1007/s10853-020-05509-2} {\bibfield  {journal} {\bibinfo
  {journal} {J. Mater. Sci.}\ }\textbf {\bibinfo {volume} {56}},\ \bibinfo
  {pages} {5321} (\bibinfo {year} {2021})}\BibitemShut {NoStop}%
\bibitem [{\citenamefont {Hosono}\ \emph {et~al.}(2015)\citenamefont {Hosono},
  \citenamefont {Tanabe}, \citenamefont {Takayama-Muromachi}, \citenamefont
  {Kageyama}, \citenamefont {Yamanaka}, \citenamefont {Kumakura}, \citenamefont
  {Nohara}, \citenamefont {Hiramatsu},\ and\ \citenamefont
  {Fujitsu}}]{10.1088/1468-6996/16/3/033503}%
  \BibitemOpen
  \bibfield  {author} {\bibinfo {author} {\bibfnamefont {H.}~\bibnamefont
  {Hosono}}, \bibinfo {author} {\bibfnamefont {K.}~\bibnamefont {Tanabe}},
  \bibinfo {author} {\bibfnamefont {E.}~\bibnamefont {Takayama-Muromachi}},
  \bibinfo {author} {\bibfnamefont {H.}~\bibnamefont {Kageyama}}, \bibinfo
  {author} {\bibfnamefont {S.}~\bibnamefont {Yamanaka}}, \bibinfo {author}
  {\bibfnamefont {H.}~\bibnamefont {Kumakura}}, \bibinfo {author}
  {\bibfnamefont {M.}~\bibnamefont {Nohara}}, \bibinfo {author} {\bibfnamefont
  {H.}~\bibnamefont {Hiramatsu}},\ and\ \bibinfo {author} {\bibfnamefont
  {S.}~\bibnamefont {Fujitsu}},\ }\bibfield  {title} {\bibinfo {title}
  {Exploration of new superconductors and functional materials, and fabrication
  of superconducting tapes and wires of iron pnictides},\ }\href
  {https://doi.org/10.1088/1468-6996/16/3/033503} {\bibfield  {journal}
  {\bibinfo  {journal} {Sci. Technol. Adv. Mater.}\ }\textbf {\bibinfo {volume}
  {16}},\ \bibinfo {pages} {033503} (\bibinfo {year} {2015})}\BibitemShut
  {NoStop}%
\bibitem [{\citenamefont {Tuleushev}\ \emph {et~al.}(2003)\citenamefont
  {Tuleushev}, \citenamefont {Volodin},\ and\ \citenamefont
  {Tuleushev}}]{10.1134/1.1633313}%
  \BibitemOpen
  \bibfield  {author} {\bibinfo {author} {\bibfnamefont {A.~Z.}\ \bibnamefont
  {Tuleushev}}, \bibinfo {author} {\bibfnamefont {V.~N.}\ \bibnamefont
  {Volodin}},\ and\ \bibinfo {author} {\bibfnamefont {Y.~Z.}\ \bibnamefont
  {Tuleushev}},\ }\bibfield  {title} {\bibinfo {title} {Novel superconducting
  niobium beryllide nb3be with a15 structure},\ }\href
  {https://doi.org/10.1134/1.1633313} {\bibfield  {journal} {\bibinfo
  {journal} {J. Exp. Theor. Phys}\ }\textbf {\bibinfo {volume} {78}},\ \bibinfo
  {pages} {440} (\bibinfo {year} {2003})}\BibitemShut {NoStop}%
\bibitem [{\citenamefont {Borlido}\ \emph {et~al.}(2022)\citenamefont
  {Borlido}, \citenamefont {Schmidt}, \citenamefont {Wang}, \citenamefont
  {Botti},\ and\ \citenamefont {Marques}}]{defpot}%
  \BibitemOpen
  \bibfield  {author} {\bibinfo {author} {\bibfnamefont {P.}~\bibnamefont
  {Borlido}}, \bibinfo {author} {\bibfnamefont {J.}~\bibnamefont {Schmidt}},
  \bibinfo {author} {\bibfnamefont {H.-C.}\ \bibnamefont {Wang}}, \bibinfo
  {author} {\bibfnamefont {S.}~\bibnamefont {Botti}},\ and\ \bibinfo {author}
  {\bibfnamefont {M.~A.~L.}\ \bibnamefont {Marques}},\ }\bibfield  {title}
  {\bibinfo {title} {Computational screening of materials with extreme gap
  deformation potentials},\ }\href {https://doi.org/10.1038/s41524-022-00811-w}
  {\bibfield  {journal} {\bibinfo  {journal} {Npj Comput. Mater.}\ }\textbf
  {\bibinfo {volume} {8}},\ \bibinfo {pages} {156} (\bibinfo {year}
  {2022})}\BibitemShut {NoStop}%
\bibitem [{\citenamefont {Arnold}\ \emph {et~al.}(1988)\citenamefont {Arnold},
  \citenamefont {Guyonnet}, \citenamefont {Giomataris}, \citenamefont
  {Pétroff}, \citenamefont {Séguinot}, \citenamefont {Tocqueville},\ and\
  \citenamefont {Ypsilantis}}]{arnoldNIaMiPRSAASDaAE1988}%
  \BibitemOpen
  \bibfield  {author} {\bibinfo {author} {\bibfnamefont {R.}~\bibnamefont
  {Arnold}}, \bibinfo {author} {\bibfnamefont {J.}~\bibnamefont {Guyonnet}},
  \bibinfo {author} {\bibfnamefont {Y.}~\bibnamefont {Giomataris}}, \bibinfo
  {author} {\bibfnamefont {P.}~\bibnamefont {Pétroff}}, \bibinfo {author}
  {\bibfnamefont {J.}~\bibnamefont {Séguinot}}, \bibinfo {author}
  {\bibfnamefont {J.}~\bibnamefont {Tocqueville}},\ and\ \bibinfo {author}
  {\bibfnamefont {T.}~\bibnamefont {Ypsilantis}},\ }\bibfield  {title}
  {\bibinfo {title} {A rich detector with a sodium fluoride radiator:
  identification up to 3 {GeV}/c},\ }\href
  {https://doi.org/10.1016/0168-9002(88)90037-X} {\bibfield  {journal}
  {\bibinfo  {journal} {Nucl. Instrum. Methods Phys. Res. A}\ }\textbf
  {\bibinfo {volume} {273}},\ \bibinfo {pages} {466} (\bibinfo {year}
  {1988})}\BibitemShut {NoStop}%
\bibitem [{\citenamefont {Frandsen}\ \emph {et~al.}(2020)\citenamefont
  {Frandsen}, \citenamefont {Nickerson}, \citenamefont {Clark}, \citenamefont
  {Solano}, \citenamefont {Baral}, \citenamefont {Williams}, \citenamefont
  {Neuefeind},\ and\ \citenamefont {Memmott}}]{frandsenJoNM2020}%
  \BibitemOpen
  \bibfield  {author} {\bibinfo {author} {\bibfnamefont {B.~A.}\ \bibnamefont
  {Frandsen}}, \bibinfo {author} {\bibfnamefont {S.~D.}\ \bibnamefont
  {Nickerson}}, \bibinfo {author} {\bibfnamefont {A.~D.}\ \bibnamefont
  {Clark}}, \bibinfo {author} {\bibfnamefont {A.}~\bibnamefont {Solano}},
  \bibinfo {author} {\bibfnamefont {R.}~\bibnamefont {Baral}}, \bibinfo
  {author} {\bibfnamefont {J.}~\bibnamefont {Williams}}, \bibinfo {author}
  {\bibfnamefont {J.}~\bibnamefont {Neuefeind}},\ and\ \bibinfo {author}
  {\bibfnamefont {M.}~\bibnamefont {Memmott}},\ }\bibfield  {title} {\bibinfo
  {title} {The structure of molten {FLiNaK}},\ }\href
  {https://doi.org/10.1016/j.jnucmat.2020.152219} {\bibfield  {journal}
  {\bibinfo  {journal} {J. Nucl. Mater.}\ }\textbf {\bibinfo {volume} {537}},\
  \bibinfo {pages} {152219} (\bibinfo {year} {2020})}\BibitemShut {NoStop}%
\bibitem [{\citenamefont {{haj Hassan}}\ \emph {et~al.}(2004)\citenamefont
  {{haj Hassan}}, \citenamefont {Akbarzadeh}, \citenamefont {Hashemifar},\ and\
  \citenamefont {Mokhtari}}]{hajhassanJoPaCoS2004}%
  \BibitemOpen
  \bibfield  {author} {\bibinfo {author} {\bibfnamefont {F.~E.}\ \bibnamefont
  {{haj Hassan}}}, \bibinfo {author} {\bibfnamefont {H.}~\bibnamefont
  {Akbarzadeh}}, \bibinfo {author} {\bibfnamefont {S.}~\bibnamefont
  {Hashemifar}},\ and\ \bibinfo {author} {\bibfnamefont {A.}~\bibnamefont
  {Mokhtari}},\ }\bibfield  {title} {\bibinfo {title} {Structural and
  electronic properties of matlockite {MFX} ({MSr}, {Ba}, {Pb}; {XCl}, {Br},
  {I}) compounds},\ }\href
  {https://doi.org/https://doi.org/10.1016/j.jpcs.2004.07.002} {\bibfield
  {journal} {\bibinfo  {journal} {J. Phys. Chem. Solids}\ }\textbf {\bibinfo
  {volume} {65}},\ \bibinfo {pages} {1871} (\bibinfo {year}
  {2004})}\BibitemShut {NoStop}%
\bibitem [{\citenamefont {Reshak}\ \emph {et~al.}(2008)\citenamefont {Reshak},
  \citenamefont {Charifi},\ and\ \citenamefont {Baaziz}}]{reshakPBCM2008}%
  \BibitemOpen
  \bibfield  {author} {\bibinfo {author} {\bibfnamefont {A.~H.}\ \bibnamefont
  {Reshak}}, \bibinfo {author} {\bibfnamefont {Z.}~\bibnamefont {Charifi}},\
  and\ \bibinfo {author} {\bibfnamefont {H.}~\bibnamefont {Baaziz}},\
  }\bibfield  {title} {\bibinfo {title} {Optical properties of the
  alkaline-earth fluorohalides matlockite-type structure {SrFX} ({X=Cl}, {Br},
  {I}) compounds},\ }\href
  {https://doi.org/https://doi.org/10.1016/j.physb.2007.09.094} {\bibfield
  {journal} {\bibinfo  {journal} {Phys. B: Condens. Matter.}\ }\textbf
  {\bibinfo {volume} {403}},\ \bibinfo {pages} {711} (\bibinfo {year}
  {2008})}\BibitemShut {NoStop}%
\bibitem [{\citenamefont {Kresse}\ and\ \citenamefont
  {Furthmüller}(1996)}]{vasp1}%
  \BibitemOpen
  \bibfield  {author} {\bibinfo {author} {\bibfnamefont {G.}~\bibnamefont
  {Kresse}}\ and\ \bibinfo {author} {\bibfnamefont {J.}~\bibnamefont
  {Furthmüller}},\ }\bibfield  {title} {\bibinfo {title} {Efficiency of
  ab-initio total energy calculations for metals and semiconductors using a
  plane-wave basis set},\ }\href
  {https://doi.org/http://dx.doi.org/10.1016/0927-0256(96)00008-0} {\bibfield
  {journal} {\bibinfo  {journal} {Comput. Mater. Sci.}\ }\textbf {\bibinfo
  {volume} {6}},\ \bibinfo {pages} {15 } (\bibinfo {year} {1996})}\BibitemShut
  {NoStop}%
\bibitem [{\citenamefont {Kresse}\ and\ \citenamefont
  {Furthm\"uller}(1996)}]{vasp2}%
  \BibitemOpen
  \bibfield  {author} {\bibinfo {author} {\bibfnamefont {G.}~\bibnamefont
  {Kresse}}\ and\ \bibinfo {author} {\bibfnamefont {J.}~\bibnamefont
  {Furthm\"uller}},\ }\bibfield  {title} {\bibinfo {title} {Efficient iterative
  schemes for ab initio total-energy calculations using a plane-wave basis
  set},\ }\href {https://doi.org/10.1103/PhysRevB.54.11169} {\bibfield
  {journal} {\bibinfo  {journal} {Phys. Rev. B}\ }\textbf {\bibinfo {volume}
  {54}},\ \bibinfo {pages} {11169} (\bibinfo {year} {1996})}\BibitemShut
  {NoStop}%
\bibitem [{\citenamefont {Bl\"ochl}(1994)}]{paw}%
  \BibitemOpen
  \bibfield  {author} {\bibinfo {author} {\bibfnamefont {P.~E.}\ \bibnamefont
  {Bl\"ochl}},\ }\bibfield  {title} {\bibinfo {title} {Projector augmented-wave
  method},\ }\href {https://doi.org/10.1103/PhysRevB.50.17953} {\bibfield
  {journal} {\bibinfo  {journal} {Phys. Rev. B}\ }\textbf {\bibinfo {volume}
  {50}},\ \bibinfo {pages} {17953} (\bibinfo {year} {1994})}\BibitemShut
  {NoStop}%
\bibitem [{\citenamefont {Kresse}\ and\ \citenamefont {Joubert}(1999)}]{paw2}%
  \BibitemOpen
  \bibfield  {author} {\bibinfo {author} {\bibfnamefont {G.}~\bibnamefont
  {Kresse}}\ and\ \bibinfo {author} {\bibfnamefont {D.}~\bibnamefont
  {Joubert}},\ }\bibfield  {title} {\bibinfo {title} {From ultrasoft
  pseudopotentials to the projector augmented-wave method},\ }\href
  {https://doi.org/10.1103/PhysRevB.59.1758} {\bibfield  {journal} {\bibinfo
  {journal} {Phys. Rev. B}\ }\textbf {\bibinfo {volume} {59}},\ \bibinfo
  {pages} {1758} (\bibinfo {year} {1999})}\BibitemShut {NoStop}%
\bibitem [{\citenamefont {Perdew}\ \emph {et~al.}(1996)\citenamefont {Perdew},
  \citenamefont {Burke},\ and\ \citenamefont {Ernzerhof}}]{PBE}%
  \BibitemOpen
  \bibfield  {author} {\bibinfo {author} {\bibfnamefont {J.~P.}\ \bibnamefont
  {Perdew}}, \bibinfo {author} {\bibfnamefont {K.}~\bibnamefont {Burke}},\ and\
  \bibinfo {author} {\bibfnamefont {M.}~\bibnamefont {Ernzerhof}},\ }\bibfield
  {title} {\bibinfo {title} {Generalized gradient approximation made simple},\
  }\href {https://doi.org/10.1103/PhysRevLett.77.3865} {\bibfield  {journal}
  {\bibinfo  {journal} {Phys. Rev. Lett.}\ }\textbf {\bibinfo {volume} {77}},\
  \bibinfo {pages} {3865} (\bibinfo {year} {1996})}\BibitemShut {NoStop}%
\bibitem [{\citenamefont {Mathew}\ \emph {et~al.}(2017)\citenamefont {Mathew},
  \citenamefont {Montoya}, \citenamefont {Faghaninia}, \citenamefont
  {Dwarakanath}, \citenamefont {Aykol}, \citenamefont {Tang}, \citenamefont
  {heng Chu}, \citenamefont {Smidt}, \citenamefont {Bocklund}, \citenamefont
  {Horton}, \citenamefont {Dagdelen}, \citenamefont {Wood}, \citenamefont
  {Liu}, \citenamefont {Neaton}, \citenamefont {Ong}, \citenamefont {Persson},\
  and\ \citenamefont {Jain}}]{atomate2017CompMatSci}%
  \BibitemOpen
  \bibfield  {author} {\bibinfo {author} {\bibfnamefont {K.}~\bibnamefont
  {Mathew}}, \bibinfo {author} {\bibfnamefont {J.~H.}\ \bibnamefont {Montoya}},
  \bibinfo {author} {\bibfnamefont {A.}~\bibnamefont {Faghaninia}}, \bibinfo
  {author} {\bibfnamefont {S.}~\bibnamefont {Dwarakanath}}, \bibinfo {author}
  {\bibfnamefont {M.}~\bibnamefont {Aykol}}, \bibinfo {author} {\bibfnamefont
  {H.}~\bibnamefont {Tang}}, \bibinfo {author} {\bibfnamefont {I.}~\bibnamefont
  {heng Chu}}, \bibinfo {author} {\bibfnamefont {T.}~\bibnamefont {Smidt}},
  \bibinfo {author} {\bibfnamefont {B.}~\bibnamefont {Bocklund}}, \bibinfo
  {author} {\bibfnamefont {M.}~\bibnamefont {Horton}}, \bibinfo {author}
  {\bibfnamefont {J.}~\bibnamefont {Dagdelen}}, \bibinfo {author}
  {\bibfnamefont {B.}~\bibnamefont {Wood}}, \bibinfo {author} {\bibfnamefont
  {Z.-K.}\ \bibnamefont {Liu}}, \bibinfo {author} {\bibfnamefont
  {J.}~\bibnamefont {Neaton}}, \bibinfo {author} {\bibfnamefont {S.~P.}\
  \bibnamefont {Ong}}, \bibinfo {author} {\bibfnamefont {K.}~\bibnamefont
  {Persson}},\ and\ \bibinfo {author} {\bibfnamefont {A.}~\bibnamefont
  {Jain}},\ }\bibfield  {title} {\bibinfo {title} {Atomate: A high-level
  interface to generate, execute, and analyze computational materials science
  workflows},\ }\href
  {https://doi.org/https://doi.org/10.1016/j.commatsci.2017.07.030} {\bibfield
  {journal} {\bibinfo  {journal} {Comput. Mater. Sci.}\ }\textbf {\bibinfo
  {volume} {139}},\ \bibinfo {pages} {140} (\bibinfo {year}
  {2017})}\BibitemShut {NoStop}%
\bibitem [{\citenamefont {de~Jong}\ \emph {et~al.}(2015)\citenamefont
  {de~Jong}, \citenamefont {Chen}, \citenamefont {Angsten}, \citenamefont
  {Jain}, \citenamefont {Notestine}, \citenamefont {Gamst}, \citenamefont
  {Sluiter}, \citenamefont {Ande}, \citenamefont {van~der Zwaag}, \citenamefont
  {Plata}, \citenamefont {Toher}, \citenamefont {Curtarolo}, \citenamefont
  {Ceder}, \citenamefont {Persson},\ and\ \citenamefont
  {Asta}}]{de2015charting}%
  \BibitemOpen
  \bibfield  {author} {\bibinfo {author} {\bibfnamefont {M.}~\bibnamefont
  {de~Jong}}, \bibinfo {author} {\bibfnamefont {W.}~\bibnamefont {Chen}},
  \bibinfo {author} {\bibfnamefont {T.}~\bibnamefont {Angsten}}, \bibinfo
  {author} {\bibfnamefont {A.}~\bibnamefont {Jain}}, \bibinfo {author}
  {\bibfnamefont {R.}~\bibnamefont {Notestine}}, \bibinfo {author}
  {\bibfnamefont {A.}~\bibnamefont {Gamst}}, \bibinfo {author} {\bibfnamefont
  {M.}~\bibnamefont {Sluiter}}, \bibinfo {author} {\bibfnamefont {C.~K.}\
  \bibnamefont {Ande}}, \bibinfo {author} {\bibfnamefont {S.}~\bibnamefont
  {van~der Zwaag}}, \bibinfo {author} {\bibfnamefont {J.~J.}\ \bibnamefont
  {Plata}}, \bibinfo {author} {\bibfnamefont {C.}~\bibnamefont {Toher}},
  \bibinfo {author} {\bibfnamefont {S.}~\bibnamefont {Curtarolo}}, \bibinfo
  {author} {\bibfnamefont {G.}~\bibnamefont {Ceder}}, \bibinfo {author}
  {\bibfnamefont {K.~A.}\ \bibnamefont {Persson}},\ and\ \bibinfo {author}
  {\bibfnamefont {M.}~\bibnamefont {Asta}},\ }\bibfield  {title} {\bibinfo
  {title} {Charting the complete elastic properties of inorganic crystalline
  compounds},\ }\href {https://doi.org/10.1038/sdata.2015.9} {\bibfield
  {journal} {\bibinfo  {journal} {Sci. Data}\ }\textbf {\bibinfo {volume}
  {2}},\ \bibinfo {pages} {150009} (\bibinfo {year} {2015})}\BibitemShut
  {NoStop}%
\bibitem [{\citenamefont {Giannozzi}\ \emph {et~al.}(2017)\citenamefont
  {Giannozzi}, \citenamefont {Andreussi}, \citenamefont {Brumme}, \citenamefont
  {Bunau}, \citenamefont {Nardelli}, \citenamefont {Calandra}, \citenamefont
  {Car}, \citenamefont {Cavazzoni}, \citenamefont {Ceresoli}, \citenamefont
  {Cococcioni}, \citenamefont {Colonna}, \citenamefont {Carnimeo},
  \citenamefont {Corso}, \citenamefont {de~Gironcoli}, \citenamefont {Delugas},
  \citenamefont {Jr}, \citenamefont {Ferretti}, \citenamefont {Floris},
  \citenamefont {Fratesi}, \citenamefont {Fugallo}, \citenamefont {Gebauer},
  \citenamefont {Gerstmann}, \citenamefont {Giustino}, \citenamefont {Gorni},
  \citenamefont {Jia}, \citenamefont {Kawamura}, \citenamefont {Ko},
  \citenamefont {Kokalj}, \citenamefont {Küçükbenli}, \citenamefont
  {Lazzeri}, \citenamefont {Marsili}, \citenamefont {Marzari}, \citenamefont
  {Mauri}, \citenamefont {Nguyen}, \citenamefont {Nguyen}, \citenamefont {de-la
  Roza}, \citenamefont {Paulatto}, \citenamefont {Poncé}, \citenamefont
  {Rocca}, \citenamefont {Sabatini}, \citenamefont {Santra}, \citenamefont
  {Schlipf}, \citenamefont {Seitsonen}, \citenamefont {Smogunov}, \citenamefont
  {Timrov}, \citenamefont {Thonhauser}, \citenamefont {Umari}, \citenamefont
  {Vast}, \citenamefont {Wu},\ and\ \citenamefont {Baroni}}]{quantumespresso}%
  \BibitemOpen
  \bibfield  {author} {\bibinfo {author} {\bibfnamefont {P.}~\bibnamefont
  {Giannozzi}}, \bibinfo {author} {\bibfnamefont {O.}~\bibnamefont
  {Andreussi}}, \bibinfo {author} {\bibfnamefont {T.}~\bibnamefont {Brumme}},
  \bibinfo {author} {\bibfnamefont {O.}~\bibnamefont {Bunau}}, \bibinfo
  {author} {\bibfnamefont {M.~B.}\ \bibnamefont {Nardelli}}, \bibinfo {author}
  {\bibfnamefont {M.}~\bibnamefont {Calandra}}, \bibinfo {author}
  {\bibfnamefont {R.}~\bibnamefont {Car}}, \bibinfo {author} {\bibfnamefont
  {C.}~\bibnamefont {Cavazzoni}}, \bibinfo {author} {\bibfnamefont
  {D.}~\bibnamefont {Ceresoli}}, \bibinfo {author} {\bibfnamefont
  {M.}~\bibnamefont {Cococcioni}}, \bibinfo {author} {\bibfnamefont
  {N.}~\bibnamefont {Colonna}}, \bibinfo {author} {\bibfnamefont
  {I.}~\bibnamefont {Carnimeo}}, \bibinfo {author} {\bibfnamefont {A.~D.}\
  \bibnamefont {Corso}}, \bibinfo {author} {\bibfnamefont {S.}~\bibnamefont
  {de~Gironcoli}}, \bibinfo {author} {\bibfnamefont {P.}~\bibnamefont
  {Delugas}}, \bibinfo {author} {\bibfnamefont {R.~A.~D.}\ \bibnamefont {Jr}},
  \bibinfo {author} {\bibfnamefont {A.}~\bibnamefont {Ferretti}}, \bibinfo
  {author} {\bibfnamefont {A.}~\bibnamefont {Floris}}, \bibinfo {author}
  {\bibfnamefont {G.}~\bibnamefont {Fratesi}}, \bibinfo {author} {\bibfnamefont
  {G.}~\bibnamefont {Fugallo}}, \bibinfo {author} {\bibfnamefont
  {R.}~\bibnamefont {Gebauer}}, \bibinfo {author} {\bibfnamefont
  {U.}~\bibnamefont {Gerstmann}}, \bibinfo {author} {\bibfnamefont
  {F.}~\bibnamefont {Giustino}}, \bibinfo {author} {\bibfnamefont
  {T.}~\bibnamefont {Gorni}}, \bibinfo {author} {\bibfnamefont
  {J.}~\bibnamefont {Jia}}, \bibinfo {author} {\bibfnamefont {M.}~\bibnamefont
  {Kawamura}}, \bibinfo {author} {\bibfnamefont {H.-Y.}\ \bibnamefont {Ko}},
  \bibinfo {author} {\bibfnamefont {A.}~\bibnamefont {Kokalj}}, \bibinfo
  {author} {\bibfnamefont {E.}~\bibnamefont {Küçükbenli}}, \bibinfo {author}
  {\bibfnamefont {M.}~\bibnamefont {Lazzeri}}, \bibinfo {author} {\bibfnamefont
  {M.}~\bibnamefont {Marsili}}, \bibinfo {author} {\bibfnamefont
  {N.}~\bibnamefont {Marzari}}, \bibinfo {author} {\bibfnamefont
  {F.}~\bibnamefont {Mauri}}, \bibinfo {author} {\bibfnamefont {N.~L.}\
  \bibnamefont {Nguyen}}, \bibinfo {author} {\bibfnamefont {H.-V.}\
  \bibnamefont {Nguyen}}, \bibinfo {author} {\bibfnamefont {A.~O.}\
  \bibnamefont {de-la Roza}}, \bibinfo {author} {\bibfnamefont
  {L.}~\bibnamefont {Paulatto}}, \bibinfo {author} {\bibfnamefont
  {S.}~\bibnamefont {Poncé}}, \bibinfo {author} {\bibfnamefont
  {D.}~\bibnamefont {Rocca}}, \bibinfo {author} {\bibfnamefont
  {R.}~\bibnamefont {Sabatini}}, \bibinfo {author} {\bibfnamefont
  {B.}~\bibnamefont {Santra}}, \bibinfo {author} {\bibfnamefont
  {M.}~\bibnamefont {Schlipf}}, \bibinfo {author} {\bibfnamefont {A.~P.}\
  \bibnamefont {Seitsonen}}, \bibinfo {author} {\bibfnamefont {A.}~\bibnamefont
  {Smogunov}}, \bibinfo {author} {\bibfnamefont {I.}~\bibnamefont {Timrov}},
  \bibinfo {author} {\bibfnamefont {T.}~\bibnamefont {Thonhauser}}, \bibinfo
  {author} {\bibfnamefont {P.}~\bibnamefont {Umari}}, \bibinfo {author}
  {\bibfnamefont {N.}~\bibnamefont {Vast}}, \bibinfo {author} {\bibfnamefont
  {X.}~\bibnamefont {Wu}},\ and\ \bibinfo {author} {\bibfnamefont
  {S.}~\bibnamefont {Baroni}},\ }\bibfield  {title} {\bibinfo {title} {Advanced
  capabilities for materials modelling with quantum espresso},\ }\href
  {https://doi.org/10.1088/1361-648X/aa8f79} {\bibfield  {journal} {\bibinfo
  {journal} {J. Condens. Matter Phys.}\ }\textbf {\bibinfo {volume} {29}},\
  \bibinfo {pages} {465901} (\bibinfo {year} {2017})}\BibitemShut {NoStop}%
\bibitem [{\citenamefont {Perdew}\ \emph {et~al.}(2008)\citenamefont {Perdew},
  \citenamefont {Ruzsinszky}, \citenamefont {Csonka}, \citenamefont {Vydrov},
  \citenamefont {Scuseria}, \citenamefont {Constantin}, \citenamefont {Zhou},\
  and\ \citenamefont {Burke}}]{pbe-sol}%
  \BibitemOpen
  \bibfield  {author} {\bibinfo {author} {\bibfnamefont {J.~P.}\ \bibnamefont
  {Perdew}}, \bibinfo {author} {\bibfnamefont {A.}~\bibnamefont {Ruzsinszky}},
  \bibinfo {author} {\bibfnamefont {G.~I.}\ \bibnamefont {Csonka}}, \bibinfo
  {author} {\bibfnamefont {O.~A.}\ \bibnamefont {Vydrov}}, \bibinfo {author}
  {\bibfnamefont {G.~E.}\ \bibnamefont {Scuseria}}, \bibinfo {author}
  {\bibfnamefont {L.~A.}\ \bibnamefont {Constantin}}, \bibinfo {author}
  {\bibfnamefont {X.}~\bibnamefont {Zhou}},\ and\ \bibinfo {author}
  {\bibfnamefont {K.}~\bibnamefont {Burke}},\ }\bibfield  {title} {\bibinfo
  {title} {Restoring the density-gradient expansion for exchange in solids and
  surfaces},\ }\href
  {https://doi.org/https://doi.org/10.1103/PhysRevLett.100.136406} {\bibfield
  {journal} {\bibinfo  {journal} {Phys. Rev. Lett.}\ }\textbf {\bibinfo
  {volume} {100}},\ \bibinfo {pages} {136406} (\bibinfo {year}
  {2008})}\BibitemShut {NoStop}%
\bibitem [{\citenamefont {{van Setten}}\ \emph {et~al.}(2018)\citenamefont
  {{van Setten}}, \citenamefont {Giantomassi}, \citenamefont {Bousquet},
  \citenamefont {Verstraete}, \citenamefont {Hamann}, \citenamefont {Gonze},\
  and\ \citenamefont {Rignanese}}]{pseudodojo}%
  \BibitemOpen
  \bibfield  {author} {\bibinfo {author} {\bibfnamefont {M.}~\bibnamefont {{van
  Setten}}}, \bibinfo {author} {\bibfnamefont {M.}~\bibnamefont {Giantomassi}},
  \bibinfo {author} {\bibfnamefont {E.}~\bibnamefont {Bousquet}}, \bibinfo
  {author} {\bibfnamefont {M.}~\bibnamefont {Verstraete}}, \bibinfo {author}
  {\bibfnamefont {D.}~\bibnamefont {Hamann}}, \bibinfo {author} {\bibfnamefont
  {X.}~\bibnamefont {Gonze}},\ and\ \bibinfo {author} {\bibfnamefont {G.-M.}\
  \bibnamefont {Rignanese}},\ }\bibfield  {title} {\bibinfo {title} {The
  pseudodojo: Training and grading a 85 element optimized norm-conserving
  pseudopotential table},\ }\href
  {https://doi.org/https://doi.org/10.1016/j.cpc.2018.01.012} {\bibfield
  {journal} {\bibinfo  {journal} {Comput. Phys. Commun.}\ }\textbf {\bibinfo
  {volume} {226}},\ \bibinfo {pages} {39} (\bibinfo {year} {2018})}\BibitemShut
  {NoStop}%
\bibitem [{\citenamefont {Lejaeghere}\ \emph {et~al.}(2014)\citenamefont
  {Lejaeghere}, \citenamefont {Speybroeck}, \citenamefont {Oost},\ and\
  \citenamefont {Cottenier}}]{eessdft}%
  \BibitemOpen
  \bibfield  {author} {\bibinfo {author} {\bibfnamefont {K.}~\bibnamefont
  {Lejaeghere}}, \bibinfo {author} {\bibfnamefont {V.~V.}\ \bibnamefont
  {Speybroeck}}, \bibinfo {author} {\bibfnamefont {G.~V.}\ \bibnamefont
  {Oost}},\ and\ \bibinfo {author} {\bibfnamefont {S.}~\bibnamefont
  {Cottenier}},\ }\bibfield  {title} {\bibinfo {title} {Error estimates for
  solid-state density-functional theory predictions: An overview by means of
  the ground-state elemental crystals},\ }\href
  {https://doi.org/10.1080/10408436.2013.772503} {\bibfield  {journal}
  {\bibinfo  {journal} {Crit. Rev. Solid State}\ }\textbf {\bibinfo {volume}
  {39}},\ \bibinfo {pages} {1} (\bibinfo {year} {2014})}\BibitemShut {NoStop}%
\bibitem [{\citenamefont {Jollet}\ \emph {et~al.}(2014)\citenamefont {Jollet},
  \citenamefont {Torrent},\ and\ \citenamefont {Holzwarth}}]{deltagauge_prime}%
  \BibitemOpen
  \bibfield  {author} {\bibinfo {author} {\bibfnamefont {F.}~\bibnamefont
  {Jollet}}, \bibinfo {author} {\bibfnamefont {M.}~\bibnamefont {Torrent}},\
  and\ \bibinfo {author} {\bibfnamefont {N.}~\bibnamefont {Holzwarth}},\
  }\bibfield  {title} {\bibinfo {title} {Generation of projector augmented-wave
  atomic data: A 71 element validated table in the xml format},\ }\href
  {https://doi.org/https://doi.org/10.1016/j.cpc.2013.12.023} {\bibfield
  {journal} {\bibinfo  {journal} {Comput. Phys. Commun.}\ }\textbf {\bibinfo
  {volume} {185}},\ \bibinfo {pages} {1246} (\bibinfo {year}
  {2014})}\BibitemShut {NoStop}%
\bibitem [{\citenamefont {Garrity}\ \emph {et~al.}(2014)\citenamefont
  {Garrity}, \citenamefont {Bennett}, \citenamefont {Rabe},\ and\ \citenamefont
  {Vanderbilt}}]{GBRV_compound}%
  \BibitemOpen
  \bibfield  {author} {\bibinfo {author} {\bibfnamefont {K.~F.}\ \bibnamefont
  {Garrity}}, \bibinfo {author} {\bibfnamefont {J.~W.}\ \bibnamefont
  {Bennett}}, \bibinfo {author} {\bibfnamefont {K.~M.}\ \bibnamefont {Rabe}},\
  and\ \bibinfo {author} {\bibfnamefont {D.}~\bibnamefont {Vanderbilt}},\
  }\bibfield  {title} {\bibinfo {title} {Pseudopotentials for high-throughput
  dft calculations},\ }\href
  {https://doi.org/https://doi.org/10.1016/j.commatsci.2013.08.053} {\bibfield
  {journal} {\bibinfo  {journal} {Comput. Mater. Sci.}\ }\textbf {\bibinfo
  {volume} {81}},\ \bibinfo {pages} {446} (\bibinfo {year} {2014})}\BibitemShut
  {NoStop}%
\bibitem [{\citenamefont {Allen}\ and\ \citenamefont
  {Dynes}(1975)}]{allen1975PRB}%
  \BibitemOpen
  \bibfield  {author} {\bibinfo {author} {\bibfnamefont {P.~B.}\ \bibnamefont
  {Allen}}\ and\ \bibinfo {author} {\bibfnamefont {R.~C.}\ \bibnamefont
  {Dynes}},\ }\bibfield  {title} {\bibinfo {title} {Transition temperature of
  strong-coupled superconductors reanalyzed},\ }\href
  {https://doi.org/10.1103/PhysRevB.12.905} {\bibfield  {journal} {\bibinfo
  {journal} {Phys. Rev. B}\ }\textbf {\bibinfo {volume} {12}},\ \bibinfo
  {pages} {905} (\bibinfo {year} {1975})}\BibitemShut {NoStop}%
\bibitem [{\citenamefont {Setyawan}\ and\ \citenamefont
  {Curtarolo}(2010)}]{setyawan2010CompMatSci}%
  \BibitemOpen
  \bibfield  {author} {\bibinfo {author} {\bibfnamefont {W.}~\bibnamefont
  {Setyawan}}\ and\ \bibinfo {author} {\bibfnamefont {S.}~\bibnamefont
  {Curtarolo}},\ }\bibfield  {title} {\bibinfo {title} {High-throughput
  electronic band structure calculations: Challenges and tools},\ }\href
  {https://doi.org/https://doi.org/10.1016/j.commatsci.2010.05.010} {\bibfield
  {journal} {\bibinfo  {journal} {Comput. Mater. Sci.}\ }\textbf {\bibinfo
  {volume} {49}},\ \bibinfo {pages} {299} (\bibinfo {year} {2010})}\BibitemShut
  {NoStop}%
\bibitem [{\citenamefont {Vaswani}\ \emph {et~al.}(2017)\citenamefont
  {Vaswani}, \citenamefont {Shazeer}, \citenamefont {Parmar}, \citenamefont
  {Uszkoreit}, \citenamefont {Jones}, \citenamefont {Gomez}, \citenamefont
  {Kaiser},\ and\ \citenamefont {Polosukhin}}]{vaswani2017attention}%
  \BibitemOpen
  \bibfield  {author} {\bibinfo {author} {\bibfnamefont {A.}~\bibnamefont
  {Vaswani}}, \bibinfo {author} {\bibfnamefont {N.}~\bibnamefont {Shazeer}},
  \bibinfo {author} {\bibfnamefont {N.}~\bibnamefont {Parmar}}, \bibinfo
  {author} {\bibfnamefont {J.}~\bibnamefont {Uszkoreit}}, \bibinfo {author}
  {\bibfnamefont {L.}~\bibnamefont {Jones}}, \bibinfo {author} {\bibfnamefont
  {A.~N.}\ \bibnamefont {Gomez}}, \bibinfo {author} {\bibfnamefont
  {{\L}.}~\bibnamefont {Kaiser}},\ and\ \bibinfo {author} {\bibfnamefont
  {I.}~\bibnamefont {Polosukhin}},\ }\bibfield  {title} {\bibinfo {title}
  {Attention is all you need},\ }in\ \href
  {https://doi.org/10.5555/3295222.3295349} {\emph {\bibinfo {booktitle} {Adv.
  Neural Inform. Process. Syst.}}}\ (\bibinfo {year} {2017})\ pp.\ \bibinfo
  {pages} {5998--6008}\BibitemShut {NoStop}%
\bibitem [{\citenamefont {Liew}\ \emph {et~al.}(2016)\citenamefont {Liew},
  \citenamefont {Khalil-Hani},\ and\ \citenamefont {Bakhteri}}]{LIEW2016718}%
  \BibitemOpen
  \bibfield  {author} {\bibinfo {author} {\bibfnamefont {S.~S.}\ \bibnamefont
  {Liew}}, \bibinfo {author} {\bibfnamefont {M.}~\bibnamefont {Khalil-Hani}},\
  and\ \bibinfo {author} {\bibfnamefont {R.}~\bibnamefont {Bakhteri}},\
  }\bibfield  {title} {\bibinfo {title} {Bounded activation functions for
  enhanced training stability of deep neural networks on visual pattern
  recognition problems},\ }\href
  {https://doi.org/https://doi.org/10.1016/j.neucom.2016.08.037} {\bibfield
  {journal} {\bibinfo  {journal} {Neurocomputing}\ }\textbf {\bibinfo {volume}
  {216}},\ \bibinfo {pages} {718} (\bibinfo {year} {2016})}\BibitemShut
  {NoStop}%
\end{thebibliography}%


%

\end{document}